%% file: main.tex
\documentclass[letterpaper,twocolumn,10pt]{article}
\usepackage{usenix-2020-09}

\renewcommand{\textrightarrow}{$\rightarrow$}

\usepackage{tikz}
\usepackage{xcolor}
\usepackage{xspace}
\usepackage{tikz}
\usepackage{amsmath}
\usepackage{pifont}
\usepackage[english]{babel}
\usepackage{blindtext}
\usepackage{lipsum}
\usepackage{amsthm}
\usepackage[ruled,vlined,linesnumbered]{algorithm2e}
\usepackage{booktabs}
\usepackage{amssymb}
\usepackage{amsfonts}

\usepackage{filecontents}
\usepackage{xspace}
\usepackage{subfig}
\usepackage{xcolor}
\usepackage{hhline}
\usepackage{multirow}
\usepackage{soul}
\usepackage{amsthm}

\usepackage{array}
\usepackage{comment}
\usepackage{listings}
\usepackage{dsfont}
\usepackage[title]{appendix}

\usepackage{caption}
\usepackage{ltablex}

\usepackage{ifluatex}

\begin{document}

\newcommand*{\affmark}[1][*]{\textsuperscript{#1}}
\newcommand*{\affaddr}[1]{#1}
\title{Automatic and Efficient Customization of \edit{Neural Networks} for ML Applications}

\author{Yuhan Liu\affmark[$\ddag$], Chengcheng Wan\affmark[$*$], Kuntai Du\affmark[$\ddag$], Henry Hoffmann\affmark[$\ddag$], Junchen Jiang\affmark[$\ddag$], Shan Lu\affmark[$\ddag$$\dag$], Michael Maire\affmark[$\ddag$]\\
\affaddr{\affaddr{\textit{\large \affmark[$\ddag$]University of Chicago }}~~~~~~\affaddr{\textit{\large \affmark[$*$]East China Normal University}}~~~~~\affaddr{\textit{\large \affmark[$\dag$]Microsoft Research}}\\
}
}

\input{macros}
\maketitle

\begin{abstract}

\edit{ML APIs} have greatly relieved application developers of the burden to \edit{design and train their own neural network models}---classifying objects in an image can now be as simple as one line of Python code to call an API. However, \edit{these APIs offer the same pre-trained models} {\em regardless} of how their output is used by different applications. This can be suboptimal as not all ML inference errors can cause application failures, and the distinction between inference errors that can or cannot cause failures varies greatly across applications. 

To tackle this problem, we first study 77 real-world applications, which collectively use six ML APIs from two \edit{providers}, to reveal common patterns of {\em how ML API output affects applications' decision processes}. Inspired by the findings, we propose {\em \tool}, an optimization framework for ML APIs, which takes effect without changing the application source code. \tool provides application developers with a parser that automatically analyzes the application to produce an abstract of its decision process, which is then used to devise an application-specific loss function that only penalizes API output errors critical to the application. \tool uses the loss function to efficiently train \edit{a neural network model} customized for each application and deploys it to serve API invocations from the respective application \edit{via existing interface}. Compared to a baseline that selects the best-of-all commercial ML API, we show that \tool reduces incorrect application decisions by \edit{43\%}.

\end{abstract}

\tightsection{Introduction}
\input{1-introduction.tex}


\tightsection{Understanding Application Decision Process}
\input{2-empirical.tex}

\tightsection{Design of \name}
\input{3-method.tex}

\tightsection{Implementation}
\input{4-implement.tex}

\tightsection{Evaluation}
\input{5-evaluation.tex}

\tightsection{Related Work}
\input{7-related_works.tex}

\tightsection{Conclusion}
\input{8-conclusion.tex}

\bibliographystyle{plain}
\bibliography{citations} 


\clearpage
\input{9-appendix.tex}
\end{document}

%% file: macros.tex
\newcommand{\shan}[1]{}
\newcommand{\hank}[1]{ }
\newcommand{\mm}[1]{ }
\newcommand{\cc}[1]{ }
\newcommand{\jc}[1]{ }
\newcommand{\yh}[1]{ }
\newcommand{\kt}[1]{ }

\newcommand{\edit}[1]{{\color{black}{#1}}}

\newcommand{\Decision}{Decision\xspace}
\newcommand{\x}{\ensuremath{\mathbf{x}}\xspace}
\newcommand{\y}{\ensuremath{\mathbf{y}}\xspace}
\newcommand{\z}{\ensuremath{z}\xspace}
\newcommand{\X}{\ensuremath{\mathbf{X}}\xspace}
\newcommand{\TestNum}{\ensuremath{T}\xspace}
\newcommand{\APIcomp}{App \circ Filter\xspace}
\newcommand{\Filter}{F\xspace}
\newcommand{\DNN}{DNN\xspace}
\newcommand{\API}{API\xspace}
\newcommand{\Appdecision}{App\xspace}
\newcommand{\total}{\ensuremath{t}\xspace}
\newcommand{\NTarget}{\ensuremath{N}\xspace}
\newcommand{\other}{\ensuremath{O}\xspace}

\newcommand{\Rate}{IDR\xspace}

\newcommand{\T}{\ensuremath{T}\xspace}
\newcommand{\Match}{\ensuremath{M}\xspace}
\newcommand{\BeforeBreak}{\ensuremath{B}\xspace}

\newcommand{\App}{$A$\xspace}
\newcommand{\M}{\ensuremath{M}\xspace}
\newcommand{\F}{\ensuremath{\alpha}\xspace}
\newcommand{\G}{\ensuremath{\beta}\xspace}
\newcommand{\Q}{\ensuremath{\gamma}\xspace}
\newcommand{\ClassId}{\ensuremath{c}\xspace}
\newcommand{\LabelId}{\ensuremath{l}\xspace}
\newcommand{\InputId}{\ensuremath{i}\xspace}
\newcommand{\TraverseId}{\ensuremath{j}\xspace}
\newcommand{\score}{effective score\xspace}
\newcommand{\scores}{effective scores\xspace}
\newcommand{\Tg}{\ensuremath{TC}\xspace}
\newcommand{\TargetSet}{\ensuremath{G}\xspace}
\newcommand{\MaxScore}{\ensuremath{f(\y)}\xspace}

\newcommand{\summary}{decision-process summary\xspace}
\newcommand{\Summary}{Decision-process summary\xspace}
\newcommand{\summaries}{decision-process summaries\xspace}
\newcommand{\target}{target class\xspace}
\newcommand{\targets}{target classes\xspace}
\newcommand{\Target}{Target class\xspace}
\newcommand{\process}{software decision process\xspace}
\newcommand{\Process}{Software decision process\xspace}
\newcommand{\bestapi}{Best-of-all API*\xspace}
\newcommand{\prespec}{Categorized models\xspace}

\newcommand{\genmodel}{generic model\xspace}
\newcommand{\genmodels}{generic models\xspace}
\newcommand{\Genmodel}{Generic model\xspace}
\newcommand{\Genmodels}{Generic models\xspace}

\newcommand{\name}{ChameleonAPI\xspace}
\newcommand{\modeld}{$\textrm{ChameleonAPI}_{basic}$\xspace}
\newcommand{\tool}{ChameleonAPI\xspace}
\newcommand{\toolbasic}{\ensuremath{\textrm{ChameleonAPI}_{basic}}\xspace}
\newcommand{\code}[1]{{\texttt{#1}}}
\newcommand{\term}[1]{\textsf{#1}}
\newcommand{\wl}{\emph{whitelist}\xspace}
\newcommand{\wls}{\emph{whitelists}\xspace}
\newcommand{\ift}{\emph{if/then branch}\xspace}
\newcommand{\ifts}{\emph{if/then branches}\xspace}

\newcommand{\fillme}{{\bf XXX}\xspace}

\newcommand*\circled[1]{\tikz[baseline=(char.base)]{
            \node[shape=circle,fill,inner sep=2pt] (char) {\textcolor{white}{\footnotesize{#1}}};}}

\newcounter{packednmbr}
\newenvironment{packedenumerate}{\begin{list}{\thepackednmbr.}{\usecounter{packednmbr}\setlength{\itemsep}{0.5pt}\addtolength{\labelwidth}{-4pt}\setlength{\leftmargin}{2ex}\setlength{\listparindent}{\parindent}\setlength{\parsep}{1pt}\setlength{\topsep}{0pt}}}{\end{list}}
\newenvironment{packeditemize}{\begin{list}{$\bullet$}{\setlength{\itemsep}{0.5pt}\addtolength{\labelwidth}{-4pt}\setlength{\leftmargin}{2ex}\setlength{\listparindent}{\parindent}\setlength{\parsep}{1pt}\setlength{\topsep}{2pt}}}{\end{list}}
\newenvironment{packedpackeditemize}{\begin{list}{$\bullet$}{\setlength{\itemsep}{0.5pt}\addtolength{\labelwidth}{-4pt}\setlength{\leftmargin}{\labelwidth}\setlength{\listparindent}{\parindent}\setlength{\parsep}{1pt}\setlength{\topsep}{0pt}}}{\end{list}}
\newenvironment{packedtrivlist}{\begin{list}{\setlength{\itemsep}{0.2pt}\addtolength{\labelwidth}{-4pt}\setlength{\leftmargin}{\labelwidth}\setlength{\listparindent}{\parindent}\setlength{\parsep}{1pt}\setlength{\topsep}{0pt}}}{\end{list}}
\let\enumerate\packedenumerate
\let\endenumerate\endpackedenumerate
\let\itemize\packeditemize
\let\enditemize\endpackeditemize

\newcommand{\tightcaption}[1]{\vspace{-0.15cm}\caption{{\normalfont{\textit{{#1}}}}}\vspace{-0.3cm}}
\newcommand{\tightsection}[1]{\vspace{-0.2cm}\section{#1}\vspace{-0.1cm}}
\newcommand{\tightsectionstar}[1]{\vspace{-0.17cm}\section*{#1}\vspace{-0.08cm}}
\newcommand{\tightsubsection}[1]{\vspace{-0.25cm}\subsection{#1}\vspace{-0.1cm}}
\newcommand{\tightsubsubsection}[1]{\vspace{-0.01in}\subsubsection{#1}\vspace{-0.01cm}}

\newcommand{\eg}{{\it e.g.,}\xspace}
\newcommand{\ie}{{\it i.e.,}\xspace}
\newcommand{\etal}{{\it et.~al}\xspace}
\newcommand{\bigO}{\mathrm{O}}
\newcommand{\twlog}{w.l.o.g.\xspac}

\newcommand{\myparashort}[1]{\vspace{0.05cm}\noindent{\bf {#1}}~}
\newcommand{\mypara}[1]{\vspace{0.05cm}\noindent{\bf {#1}:}~}
\newcommand{\myparatight}[1]{\vspace{0.02cm}\noindent{\bf {#1}:}~}
\newcommand{\myparaq}[1]{\smallskip\noindent{\bf {#1}?}~}
\newcommand{\myparaittight}[1]{\smallskip\noindent{\emph {#1}:}~}
\newcommand{\question}[1]{\smallskip\noindent{\emph{Q:~#1}}\smallskip}
\newcommand{\myparaqtight}[1]{\smallskip\noindent{\bf {#1}}~}

\newcommand{\cmark}{\ding{51}}%
\newcommand{\xmark}{\ding{55}}%



\definecolor{backcolour}{rgb}{0.96,0.96,0.96}
\definecolor{codegray}{rgb}{0.5,0.5,0.5}
\definecolor{deepblue}{rgb}{0,0,0.6}
\definecolor{deepred}{rgb}{0.6,0,0}
\definecolor{deepgreen}{rgb}{0,0.5,0}
\lstdefinestyle{mystyle}{
    backgroundcolor=\color{backcolour},   
    commentstyle=\color{codegreen},
    morekeywords={self, True},
    keywordstyle=\color{deepblue},
    numberstyle=\tiny\color{codegray},
    emph={MyClass,__init__,EncodingType,Image},
    emphstyle=\color{deepred},
    stringstyle=\color{deepgreen},
    basicstyle=\ttfamily\footnotesize,
    breakatwhitespace=false,         
    breaklines=true,                 
    captionpos=b,                    
    keepspaces=true,                 
    numbers=left,                    
    numbersep=5pt,                  
    showspaces=false,                
    showstringspaces=false,
    showtabs=false,                  
    tabsize=1
}

%% file: 1-introduction.tex
\label{sec:intro}

The landscape of ML applications has greatly changed, with the rise of \edit{ML APIs} significantly lowering the barrier of ML application developers. \edit{Instead of 
designing and managing neural network models by themselves via frameworks like TensorFlow and PyTorch, application developers can now simply invoke ML APIs, 
provided by open-source libraries or commercial cloud service providers, to
accomplish common ML tasks like object detection, facial emotion analysis, etc}. \edit{This convenience} thus gives rise to a variety of ML applications on smartphones, tablets, sensors, and personal assistants~\cite{9778241,9453402,10.1145/3555802,wan2021machine}.

\edit{Although ML APIs have} eased the integration \edit{of ML tasks} with applications, \edit{they are} suboptimal \edit{by serving different applications
with the same neural network models}. This issue is particularly striking when applications use \edit{the ML API results} to make control-flow decisions (also referred to as \textit{application decisions} in this paper). Different applications may check the result of the same ML API using different control-flow code structures and different condition predicates, a process that we refer to as the application's {\em decision process} (see \S\ref{sec:empirical} for the formal definition). Due to the heterogeneity across applications' decision processes, we make two observations.
\begin{packeditemize}
\item First, some incorrect ML API outputs may still lead to correct application decisions, with only certain {\em critical errors} of API output affecting the application's decision.
\item  Second, among all possible output errors of an ML API, which ones are critical {\em vary} significantly across applications that use this API. That is, the same API output error may have a much {\em greater} effect on one application than on another. 
\end{packeditemize}

\begin{figure}
    \includegraphics[width=0.9\linewidth]{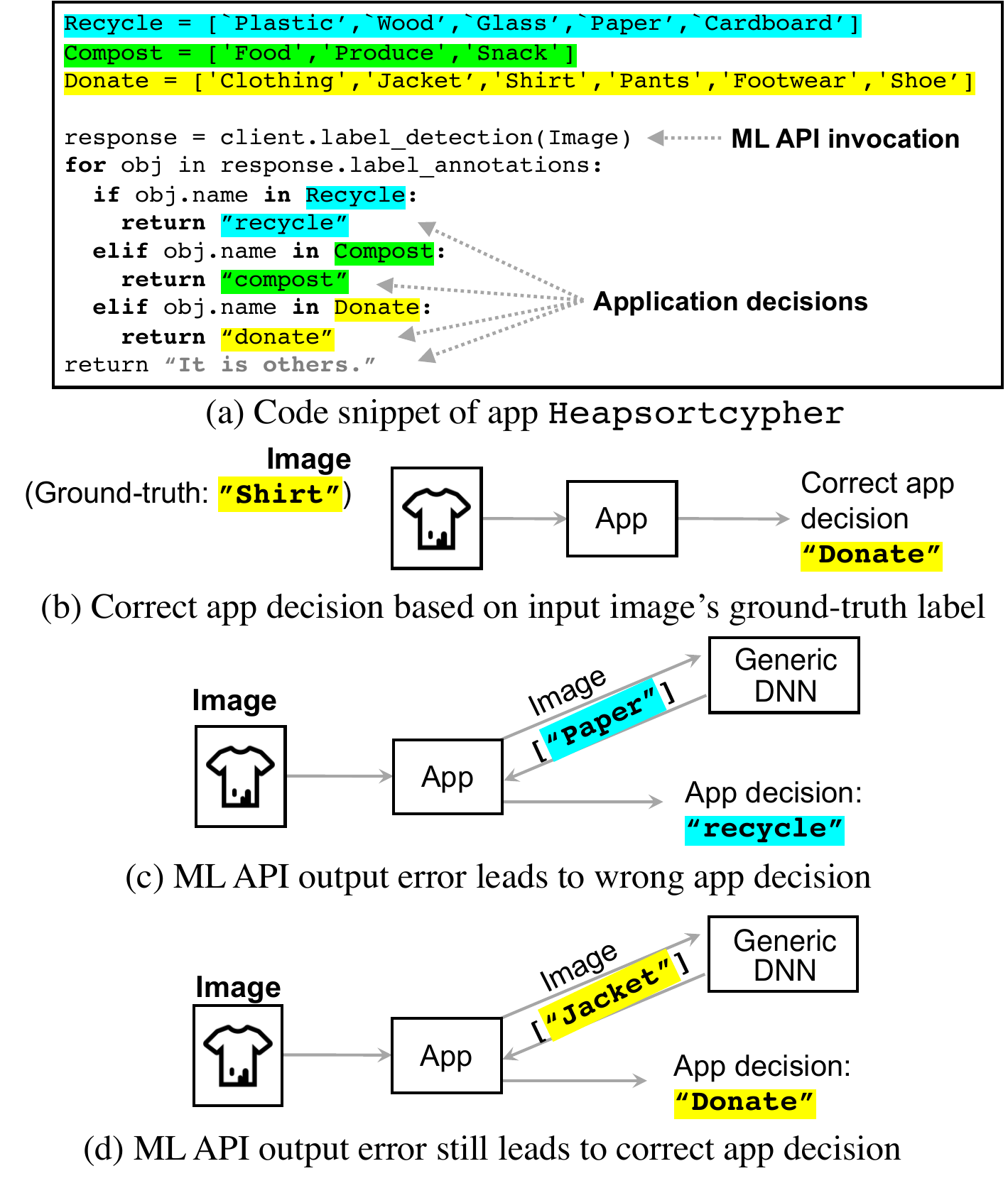}
    \tightcaption{An example ML application whose decision depends on the output of ML API (multi-label classification), but not all errors of ML API output have the same effect.
    }
    \label{fig:input_example}
\end{figure}

Figure~\ref{fig:input_example} illustrates the decision process of a garbage-classification application \code{Heapsortcypher}~\cite{heapsort}. It first \edit{invokes} Google's classification API upon a garbage image. Then, based on the returned labels, a simple logic is used to make the \emph{application decision}
about which one of the pre-defined categories (\code{Recycle}, \code{Compost}, and \code{Donate}) or \code{others} the image belongs to. 
For example, for an input image whose ground-truth label is \code{``Shirt''}, the correct application decision is \code{Donate}, as shown in Figure \ref{fig:input_example} (b).

For this application, when the classification API fails to return \code{``Shirt''}, 
the application decision may or may not be wrong. For example,
Figure~\ref{fig:input_example} (c) and (d) show two possible wrong API output: if the output is \code{``Paper''}, the application will make a wrong decision of \code{Recycle}; however, if the output is \code{``Jacket''}, the application will make the correct decision of \code{Donate} despite not matching the ground-truth label.
\edit{
More subtly, if the API returns a list of two labels, \code{``Shirt''}
and \code{``Paper''}, the application would make a correct decision if \code{``Shirt''}
is ordered before \code{``Paper''} by the API, but would make a wrong decision if
\code{``Paper''} is ordered before \code{``Shirt''}. The reason is that the application
logic, the \code{for} loop in Figure \ref{fig:input_example} (a), checks one API-output label at a time. As we will see later, there are also other ways that applications check the API-output list, which will affect application decision differently.}

As we can see, for a specific application, some errors of an ML API may be critical, like mis-classifying the shirt to \code{``Paper''} in the example above, and yet some errors may be non-critical, like mis-classifying the shirt image as \code{``Jacket''} 
\edit{or classifying the shirt image as both \code{``Shirt''} and \code{``Paper''}} in the examples above. Which errors are critical varies, depending on the application's decision process. 


These observations 
regarding the critical errors specific to each application 
suggest substantial room for improvement by customizing the ML API\edit{, essentially the neural network model underneath the API,} for individual application's decision process. 
In particular, for a given application, the \edit{customized model}
\edit{
can afford having more errors less critical to the application for the benefit of
having fewer critical errors that cause wrong application decisions.}

Thus, our goal is \edit{to allow ML APIs and their underlying neural network models to be {\em automatically} customized for a given application, so as to {\em minimize} incorrect application decisions {\em without} changing the application's source code or interface between ML API and software exposed to developers.}
This way, application developers who do not have the expertise to design and train customized ML models can still enjoy the accessibility of \edit{generic ML APIs} while getting closer to the accuracy of ML models customized for the application.

No prior work shares the same goal as us. The closest line of prior work specializes DNN models for given queries~\cite{kang2018blazeit,kang2017noscope,koudas2020video,figo,cao2021thia}, but they require application developers to use a domain specific language (\eg in SQL~\cite{kang2018blazeit}) instead of general programming languages, like Java and Python, and mostly focus on reducing the DNN's size. In contrast, we keep both the ML API interface  and the application source code intact while avoiding incorrect decisions for ML applications.

With the aforementioned goal, this paper makes two contributions.
\emph{First,} we run an empirical study over 77 real-world applications that collectively use six ML APIs to reveal several common patterns of how the outputs of ML APIs affect the application decisions (\S\ref{sec:empirical}). 

Our study identifies two types of ML API output that are used by applications to make {control-flow} decisions (categorical labels and sentiment scores), and three types of decision types (True-False, Multi-Choice, and Multi-Selection) with different implications regarding which ML API output errors are critical to the application. 

Our study also quantitatively reveals opportunities of model customization. (1) Although popular image-classification models are trained to recognize as many as  19.8K different labels, the largest number used by any one application for decision making is only 54. Consequently, mis-classification among the remaining tens of thousands of labels are completely irrelevant to an application. 
(2) More importantly, applications tend to treat multiple labels (4.7 on average) as one equivalence class in their decision making, such as labels
\code{Plastic}, \code{Wood}, \code{Glass}, \code{Paper}, and \code{Cardboard} in Figure \ref{fig:input_example}(a).
Mis-classification among those labels inside one equivalence class does not matter. (3) Which labels are relevant to an application's decision making vary greatly across applications, with only 12\% of application pairs share any labels used for their decision making.

Second, inspired by the empirical study, we propose \tool, which customizes and serves ML models behind the ML API for each given application's decision process, without any change to the existing ML API or the application source code (\S\ref{sec:method}). \tool works in three steps. First, it provides a parser that analyzes application source code to extract information about how ML inference results are used in the application's decision process. Based on the analysis result, \tool then constructs the loss function to reflect which \edit{ML model} output is more relevant to the given application as well as the different severity of ML inference errors on the application decisions. The ML model will be retrained accordingly using the new loss function. Finally, when the ML API is \edit{invoked by the application at runtime}, a customized ML model \edit{will be used to serve this query}.


\edit{We evaluate \tool on 57 real-world open-source applications that use Google and Amazon's vision and language APIs. We show that \tool's re-trained models reduce  48\% of incorrect decisions compared to the off-the-shelf ML models and 50\% compared to the commercial ML APIs.
Even compared with a baseline that selects the best-of-all commercial ML API, \tool reduces 43\% of incorrect decisions. 
\tool only takes up to \edit{24} minutes on a  GeForce RTX 3080 GPU to re-train the ML model.}

%% file: 2-empirical.tex
\label{sec:empirical}

We conduct an empirical study to understand \edit{how applications make decisions based on ML APIs} (\S\ref{subsec:study-mech}), and how this decision making logic implies the different severity of ML inference errors (\S\ref{subsec:study-imp}). This study will reveal why and how to customize the ML API backend for each application. \edit{As a representative sample of ML APIs, this study focuses on cloud AI services due to their popularity.}

\tightsubsection{Definitions}
\label{sec:definitions}

\mypara{Preliminaries} We begin with basic definitions.

\begin{packeditemize}
    \item \emph{Application decision}: the collective control-flow decisions (i.e., which branch(es) are taken) made by the application under the influence of a particular ML API output.
    \item {\em Incorrect ML API output:} a situation when the API output differs from the API input's human-labeled ground truth. We refer to such ML API outputs as {\em API output errors}.
    \item \emph{Correct decision}: the application decision if the API output is the same as the human-labeled ground-truth of the input. 
    \item \emph{Application decision failure}: a situation when the application decision is different from the correct decision, also referred to as \textit{application failure} for short in this paper.
\end{packeditemize}

\mypara{\Process}
Given these definitions, an application's {\em \process} (or decision process for short) is the logic that maps an ML API output to an application decision. 
The code snippet in Figure~\ref{fig:input_example} shows an example decision process, which maps the output of a classification ML API on an image to the image's recycling categorization specific to this application.

\mypara{Critical and non-critical errors}
For a given decision process, some API output errors will still lead to a correct decision, whereas some API output errors will lead to an incorrect decision and hence an application failure.
We refer to the former as {\em non-critical errors}, and the latter as {\em critical errors}.

\tightsubsection{Methodology}
\label{sec:methodology}
 
\begin{table}[]
\begin{footnotesize}
   \begin{tabular}{@{}lllr@{}}
\toprule
ML API name          & ML task              & Provider & \# of apps \\ \midrule
label\_detection     & Vision::Image classification & Google           & 29         \\
detect\_labels       & Vision::Image classification & Amazon           & 11         \\
object\_localization & Vision::Object detection     & Google           & 8          \\
analyze\_sentiment   & Language::Sentiment analysis   & Google           & 14         \\
analyze\_entities    & Language::Entity recognition   & Google           & 6          \\
classify\_text       & Language::Text classification  & Google           & 9          \\ \bottomrule
\end{tabular} 
\end{footnotesize}
\tightcaption{Summary of applications used in our empirical study. }
\label{tab:all_stats}
\end{table}

Our work focuses on applications that use ML API output to make control-flow decisions. 
To this end, we look at 77 open-source applications which collectively use six widely used vision and language APIs~\cite{wan2021machine, chen2021did} offered by two popular \edit{cloud AI service} providers, as summarized in Table~\ref{tab:all_stats}. 

These applications come from two sources. First, we study all  50 applications that use vision and language APIs from a recently published benchmark suite of open-source ML applications~\cite{wan2022automated}. Second, given the popularity of image classification APIs~\cite{chen2020frugalml,chen2022frugalmct}, we additionally sample 27 applications from GitHub 
that use Google and Amazon image classification APIs
(16 for the former and 11 for the latter). We obtain these 27 by checking close to 100 applications that use image classification APIs and filtering out those that directly print out or store the API output.
Every application in our benchmark suite uses exactly one ML API for decision making. 

\mypara{Threats to validity} 
While many applications use the APIs listed in Table~\ref{tab:all_stats}, there are a few other APIs not covered in our study. A few vision and language-related ML tasks are not as popular and hence
are not covered in our study (\eg face recognition and syntax analysis).
Speech APIs are not covered, because their outputs are rarely used to affect application control flow based on our checking of open-source applications. \edit{Finally, our study
does not cover applications that use ML APIs offered by other cloud or local providers. 
}

\tightsubsection{Understanding the decision mechanism}
\label{subsec:study-mech}

\begin{figure*}
\centering
     \includegraphics[width=0.99\linewidth]{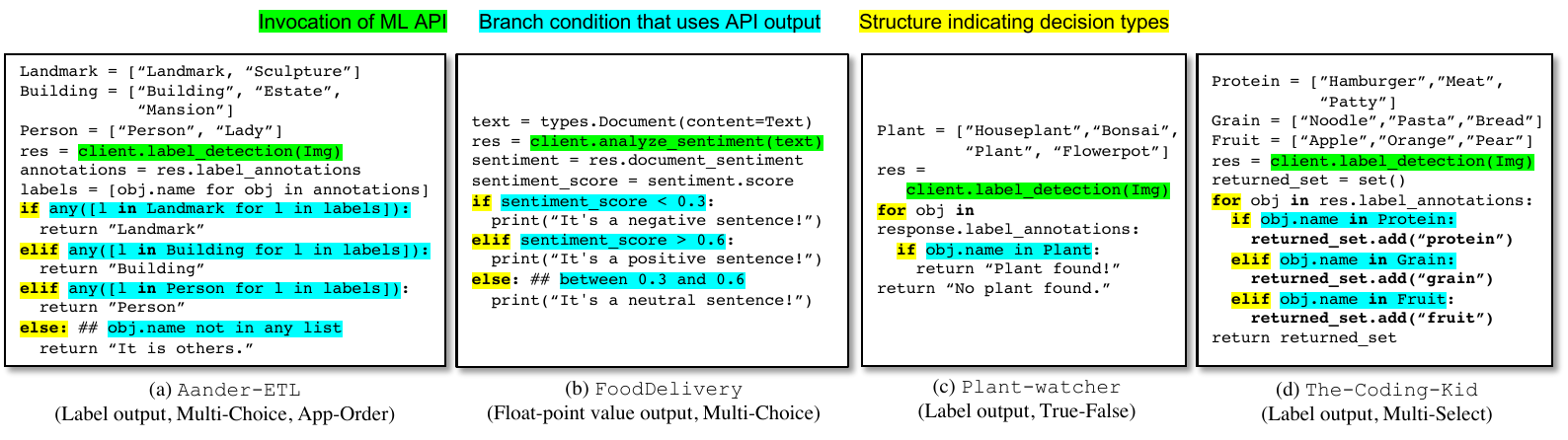}
    \tightcaption{Code snippets from five example applications where ML API output affects control flow decisions in different ways.}
\label{fig:code}
\end{figure*}

\emph{\textbf{Q1}: {What types of ML API outputs are typically used by applications to make decisions? } }

ML APIs produce output of a variety of types.
The sentiment analysis API outputs a list of floating-point value pairs (\code{score} and 
\code{magnitude}), describing the sentiment of the whole document and every individual sentence; the other five APIs in Table \ref{tab:all_stats} each 
produces a list of categorical labels ranked in {descending} order of their confidence scores, which is also part of the output. Some APIs' output also 
contains other information, like coordinates of bounding boxes, 
entity names, links to Wikipedia URLs, and so on.
Among all these, only two types have been used in application decision processes of our studied application: the floating-point pair (\code{score} and \code{magnitude})
and the categorical labels.

For the 63 applications that use categorical-label output from the five APIs (all except \code{analyze\_sentiment} in Table \ref{tab:all_stats}), they each define one or more {\em label lists} and check which label list(s) an API output label belongs to. The code snippet of a landmark classification application in Figure~\ref{fig:code}(a) is an example of this. It calls \edit{the \code{label\_detection} API} with a sight-seeing image and checks the output labels to see if the image might contain \code{Landmark}, or just ordinary \code{Building}, or \code{Person}.

For the 14 applications that use the \code{analyze\_sentiment} API, they each define several \emph{value ranges} and check which range the sentiment
\code{score} and/or \code{magnitude} falls in. 
The code snippet of  \code{FoodDelivery}~\cite{food-delivery}
in Figure~\ref{fig:code}(b) is an example.
This application calls \code{analyze\_sentiment} with a restaurant review text, and then checks the returned
sentiment \texttt{score} to judge if the review is negative, positive, or neutral.

\emph{\textbf{Q2}: What type of decisions do applications make?} 

We observe three categories of ML-based decision making, 
which we name following common question types in exams:

(1) \term{True-False} decision, where a single label list or value range
is defined and one selection is allowed: either the ML API output belongs to this  list/range or not.
This type occurs in about one third of the applications in our study. 
For example, the plant management application \code{Plant-watcher}~\cite{plant-watcher} (Figure~\ref{fig:code}(c)) checks to see if the image contains plants or not.

(2) \term{Multi-Choice} decision, where 
multiple
lists of labels or value ranges are defined, and one selection is allowed.
The ML API output will be assigned to \emph{at most one} list or range; the application's decision making logic determines which of these lists/ranges the output belongs to, or determines that the output belongs to none of them.
This type of decision is the most common, occurring in about 45\% of benchmark applications.
The garbage classification application discussed in \S\ref{sec:intro} makes such a \term{Multi-Choice} decision. 
It decides which one of the following classes the input image belongs to:
\code{Recycle}, \code{Compost}, \code{Donate}, or none of them.

(3) \term{Multi-Select} decision, where multiple label lists or value ranges are defined, and {\em multiple} selections are allowed about which label lists or value ranges the ML API output belongs to. 
This type of decisions occur in close to a quarter of the applications.
Figure~\ref{fig:code}(d) illustrates such an example
from the nutrition advisor application \code{The-Coding-Kid}~\cite{coding-kid}. 
This application defines three  label lists to represent nutrition types: \code{Protein}, \code{Grain}, and \code{Fruit}, and it checks to find all the nutrition types present in the input image.

In the remainder of the paper, we will use {\bf \target} to refer to a label list (or a value range) that is used to match against a categorical label (or a value). 
For instance, the code snippet in Figure~\ref{fig:code}(a) has three label lists as its \targets (\code{[Landmark, Sculpture]}, \code{[Building, Estate, Mansion]}, and \code{[Person, Lady]}), and the code snippet in Figure~\ref{fig:code}(b) has three value ranges as its \targets (\code{<0.3}, \code{>0.6}, and in between).

\emph{\textbf{Q3}: How do applications reach \term{Multi-Choice} decisions?
}

When the ML API outputs multiple labels, the outcome of a \term{Multi-Choice} decision varies depending on which {\em matching order} is used.
First, the matching order can be determined by the API output. 
For example, the garbage classification application
(Figure~\ref{fig:input_example})
first checks whether the first label in the API output matches any \target. 
If so, later API output labels will be skipped, even if they might match with a different class. 
If there is no match for the first label, the second output label is checked, and so on. These labels are ranked by the API in the descending order of
their associated confidence scores, so we refer to such a matching order as
\term{API-order}. It is used by 80\% of applications that make \term{Multi-Choice} decisions.

The matching order can also be specified by the application, referred to as \term{App-order}.
For instance, 
regardless the API output, application \code{Aander-ETL}~\cite{aander-etl} (Figure~\ref{fig:code}(a)) always first checks if the \code{Landmark} class matches with \emph{any} output label.
If there is a match, the decision is made. Only when it fails to match
\code{Landmark}, will it move on to check the next choice, \code{Building}, and so on.
This matching order is used by 20\% of applications that make \term{Multi-Choice} decisions.

\tightsubsection{Understanding the decision implication}
\label{subsec:study-imp}

\emph{\textbf{Q4}: Does an application need ML APIs that can accurately identify thousands of labels?}

ML models \edit{behind popular ML APIs} are well trained to \edit{support a wide range of} applications. For example, Google and Microsoft's image-classification APIs are capable of identifying more than 10000 labels \cite{openimages}, while Amazon's image-classification API can identify 2580 labels \cite{amazon-label}. However, for each individual application, its decision making only requires classifying the input image into a handful of \targets: 7 at most in our benchmark applications. The largest number of image-classification labels checked by an application is 54, a tiny portion of all the labels an image-classification API could output.

Clearly, for any application, a customized ML model that focuses on those \targets used by the application's decision process has the potentially to out-perform the big and generic ML model behind ML APIs. How to accomplish the customization without damaging the accessibility of \edit{ML APIs} will be the goal of \tool.

\emph{\textbf{Q5}: Are there equivalence classes among ML API outputs in the context of application decision making?}

For the 63 applications that make decisions based on API output of categorical labels, they present 121 \targets in total, each containing 4.7 labels on average (3 being the median). Only 35 \targets in 22 applications contain a single label.
For the 14 applications that make decisions based on floating-point sentiment \code{score} and \code{magnitude}, their \targets \emph{all} contain an infinite number of \code{score} or \code{magnitude} values. In other word, no class contains just a single value. 

Clearly, the wide presence of multi-value \targets creates equivalence classes among output returned by the API---errors within one equivalence class are \emph{not} critical to the corresponding application. This offers another opportunity for ML customization.

\emph{\textbf{Q6}: How much difference is there between different applications' \targets?} 
 
Overall, the difference is significant. We have conducted pair-wise comparison
between any two applications in our benchmark suit, and found that 88\% of 
application pairs share \emph{no} common labels in any of their \targets. 
Similarly, among the 381 labels that appear in at least one application's 
\targets, 88\% of them appear in only one application
(\ie 335 out of 381 labels).

Clearly, there is {\em little overlap} among the \targets of different applications, again making a case for {\em per-application} customization of the ML models used by the ML APIs.

\emph{\textbf{Q7}: Do different decision mechanisms imply different sensitivity to output errors of ML APIs?}

Even for two applications that have the same \targets, if they try to make
different types of decisions,
they will have different sensitivity to ML API output errors---some 
 API errors might be critical to one application, but not to the other.
For example, errors that affect the selection of different \targets are equally critical to \term{Multi-Select} decisions. 
However, this is not true for \term{Multi-Choice}, where only the first matched \target matters.
Furthermore, the matching order of a \term{Multi-Choice} decision affects which errors are critical.
When the \term{API-output} order is used (\eg \code{HeapsortCypher} in Figure~\ref{fig:input_example}), an error on the first label in the API output is more likely to be critical than an error on other labels in the output.
However, when \term{App-order} order is used (\eg \code{Aander-ETL} in Figure~\ref{fig:code}(a)), errors related to labels in the first \target (\eg \code{Landmark}) are more likely to be critical than those related to labels in later \targets (\eg \code{Person}). 

Clearly, to customize ML models for each application, we need to take into account what is the decision type and what is the matching order (for \term{Multi-Choice} decisions).

%% file: 3-method.tex
\label{sec:method}

Inspired by the study of \S\ref{sec:empirical}, we now present \tool which automatically customizes ML models for applications.

\tightsubsection{Problem formulation}
\label{sec:overall}

\begin{figure}
    \centering
    \includegraphics[width=0.99\linewidth]{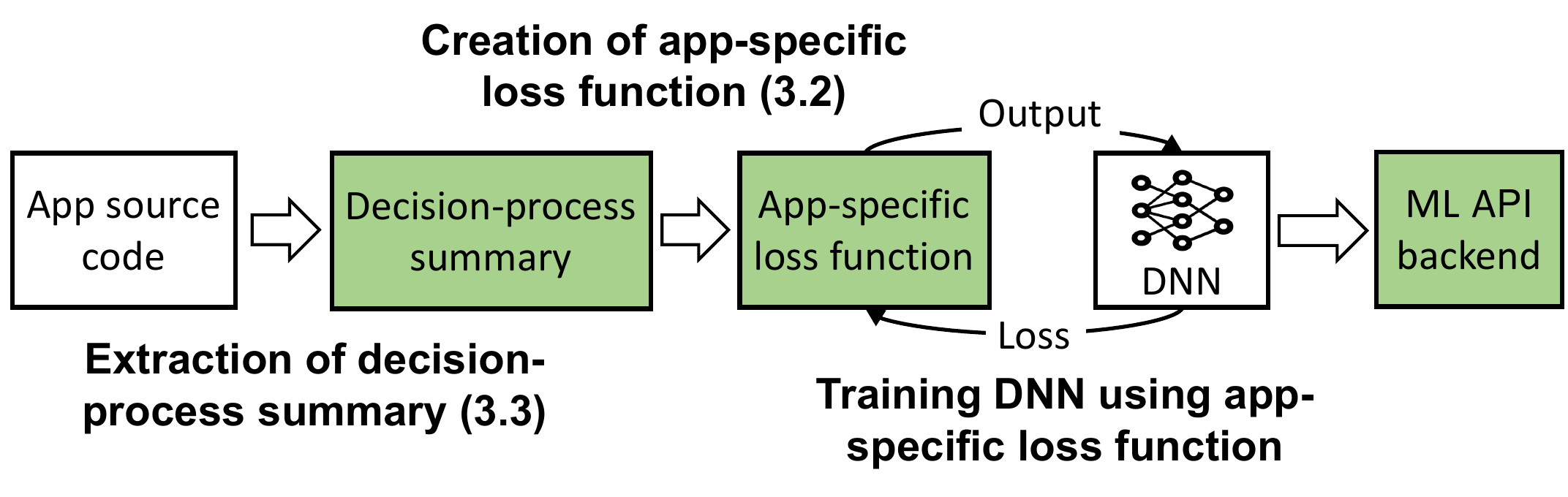}
    \tightcaption{
    The logical steps of how \tool customizes \edit{ML APIs} for individual applications.
    }
    \label{fig:steps}
\end{figure}

\mypara{Goal} For an application that uses \edit{ML APIs, our goal is to
{\bf minimize critical errors} in the API outputs for this application by efficiently re-training the original generic neural network models underneath these APIs into customized models}; our approach stands in contrast to typical approaches that minimize {\em all} inference errors. In other words, the new ML model should return outputs that lead the application process to the same decision as if the ground-truth of the input is returned by the ML API. 

To formally state this objective, we denote how an application makes a decision by $\Appdecision(\API(\x))$, where $\x$ is the input to the ML API and $\API(\x)$ is the API output. 
Then for a given application decision process of $\Appdecision(\cdot)$ and an input set $\X$\footnote{A careful reader might notice that the formulation in Eq.~\ref{eq:goal} also depends on the input set.
Though the input set should ideally follow the same distribution of real user inputs of the application, this distribution is hard to obtain in advance and may also vary over time and across users.
Instead, we focus our discussion on training the ML model to minimize Eq.~\ref{eq:goal} with an assumed input distribution. Our evaluation (\S\ref{sec:evaluation}) will test the resulting model's performance over different input distributions.}, our goal is to train an ML model $\DNN(\cdot)$ such that 
\begin{align}
    \min_{\x_\InputId \in \X} \left|\{\x_\InputId | \Appdecision(\API(\x_\InputId)) \neq \Appdecision(\widehat{\API}(\x_\InputId))\}\right|, \nonumber \\
    \textrm{where }\API(\x_\InputId)=\Filter(\DNN(\x_\InputId))
    \label{eq:goal}
\end{align}

Here, $\widehat{\API}(\x_\InputId)$ is a hypothetical API function that always returns the ground truth of input $\x_\InputId$, and $\Filter(\cdot)$ represents the postprocessing used by the API to translate a DNN output to an API output.
For instance, an image classification model's output is a vector of confidence scores between 0 and 1 (each for a label),
but the ML API will use a threshold $\theta$ to filter and return only labels with scores higher than $\theta$, or the
top $k$ labels with the highest confidence scores.

Our goal in Eq~\ref{eq:goal} differs from the traditional goal of an ML model, which minimizes any errors in the API output, \ie 
\begin{align}
\min_{\x_\InputId \in \X} \left|\{\x_\InputId | \API(\x_\InputId) \neq \widehat{\API}(\x_\InputId)\}\right|.
\label{eq:goal_old}
\end{align}
Given that it is hard to obtain a DNN with 100\% accuracy, the difference between the two formulations is crucial, since not  all API output errors in Eq.~\ref{eq:goal_old} will cause incorrect application decisions in Eq.~\ref{eq:goal}.
Thus, compared to optimizing Eq.~\ref{eq:goal_old}, optimizing Eq.~\ref{eq:goal} is more likely to focus the DNN training on reducing the critical errors for the application.

To train a DNN that optimizes Eq.~\ref{eq:goal}, \edit{we need to decide} 
if a DNN inference output $\DNN(\x)$ is a critical error or not (\ie $\Appdecision(\DNN(\x))\neq\Appdecision(\widehat{\API}(\x))$) at the end of \textit{every} training iteration. \edit{This decision needs to be made automatically
and efficiently. For example,
repeatedly running the entire ML application after every training iteration 
would not work, as it may significantly slow down the training procedure.} 

\mypara{Logical steps of \tool} To customize and deploy the DNN for an application, \tool takes three logical steps (Figure~\ref{fig:steps}). First, \tool extracts from an application's source code a \emph{\summary} (explained shortly), a succinct representation of the application's decision process, which will be used to determine if a DNN inference error is critical (details in \S\ref{sec:extract}).  Second, \tool converts a \summary to a {\em loss function}, which can be directly used to train a DNN (details in \S\ref{sec:loss}). This loss function only penalizes DNN outputs that lead to critical errors with respect to a given application.  Finally, the loss function will be used to \edit{train a customized DNN for this particular application's ML API invocations (\S\ref{sec:runtime}). }

\myparashort{A \summary}is a succinct abstraction of the application that contains enough information to determine if a DNN inference output causes a critical error or not. 
Specifically, it includes three pieces of information (defined in \S\ref{subsec:study-mech}): 
\begin{packeditemize}
\item \emph{Composition of \targets:} 
the label list or value range of each \target; 
\item {\em Decision type:} \term{True-False}, \term{Multi-Choice}, or \term{Multi-Select};
\item {\em Matching order:} over the target classes, \term{API-order} or \term{App-order}, if the application makes a \term{Multi-Choice} decision.
\end{packeditemize}
For a concrete example, the \summary of the garbage classification application in Figure~\ref{fig:code}(a) contains
(1) three label lists representing three \targets: \code{Recycle}, \code{Compost}, and \code{Donate}; 
(2) the \term{Multi-Choice} type of decision; and 
(3) the matching order of \term{API-order}.

\mypara{What is changed, what is not}
\tool does {\em not} change the ML API or the application source code. Unlike recent work that aims to shrink the size of DNNs or speed them up~\cite{mullapudi2019online,kang2018blazeit,kang2017noscope}, we do not change the DNN architecture (shape and input/output interface); instead, we train the DNN to minimize critical errors. That said, deploying \tool has two requirements. First, the application developers need to run \tool's parser script to automatically extract the \summary.
Second, an ML model \edit{needs to be retrained for each} application, instead of serving the same model to all applications.

The remainder of this section will begin with the design of the application-specific loss function based on \summary,
followed by how to extract the \summary from the application, and finally, how the customized ML models are used to serve ML API queries.

\tightsubsection{Application-specific loss function}
\label{sec:loss}

Given Eq~\ref{eq:goal}, \tool trains a DNN model with a {\em new loss function}, which only penalizes critical errors of an application, rather than all DNN inference errors.
Since decision processes vary greatly across applications (\S\ref{subsec:study-imp}), we first explain how to conceptually capture different decision processes in a generic description, which allows us to derive the mathematical form of \tool's loss function later.

\begin{figure}
\centering
     \includegraphics[width=0.94\linewidth]{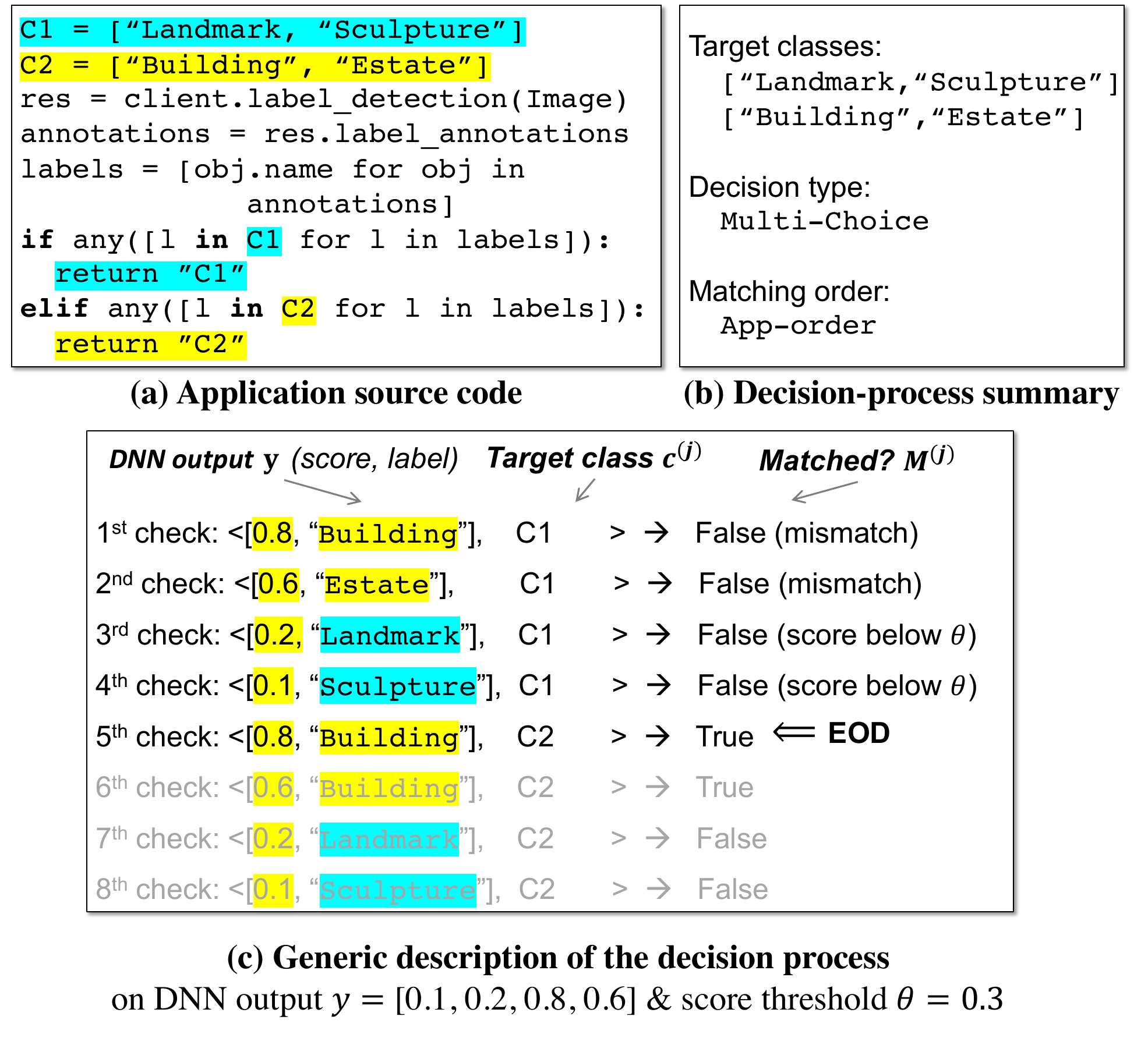}
    \tightcaption{The generic description (shown in (c)) of an application (whose source code is shown in (a) and \summary in (b)) on a DNN inference output $\y$.
}
\label{fig:loss}
\end{figure}

\mypara{Generalization of decision processes}
For each application in our study (\S\ref{sec:methodology}), our insight is that its decision process can always be viewed as traversing a sequence of conditional {\em checks} until an {\em end-of-decision} ({\em EOD}) occurs:
\vspace{-0.4cm}

{\small
\begin{align*}
1^{\textrm{st}}\textrm{ check:} &<\y,\ClassId^{(1)}>\rightarrow\Match^{(1)}&\\
&\dots&\\
\TraverseId^{\textrm{th}}\textrm{ check:} &<\y,\ClassId^{(\TraverseId)}>\rightarrow\Match^{(\TraverseId)}& \Leftarrow \textrm{\bf EOD}\\
&\dots&
\vspace{-0.4cm}
\end{align*}
}
\noindent
where the $\TraverseId$-th check takes as input the DNN output $\y$ and one of \targets $\ClassId^{(\TraverseId)}$, 
and returns a binary $\Match^{(\TraverseId)}$ indicating whether $\y^{(\TraverseId)}$ {\em matches} the condition of $\ClassId^{(\TraverseId)}$ and a binary decision whether this check happens before the EOD.
The set of \targets successfully matched before the EOD will be those selected by the application.

Figure~\ref{fig:loss} shows (a) an example application, (b) the decision-process summary,  and (c) the generic description for this application's decision process and a DNN output.  

This generic description (\eg the traversal order of the \targets, how a match is determined in a check, and when the EOD occurs) will depend on the information in the \summary and the DNN output $\y$. 
We stress that this generic description may {\em not} apply to all applications, but it does apply to all applications in our study (\S\ref{sec:methodology}).

\mypara{Categorization of critical errors}
Importantly, this generic description helps to categorize critical errors:
\begin{packeditemize}
\item {\em Type-1 Critical Errors:} A correct \target $\ClassId$ is not matched before EOD, but will be so if EOD occurs later. 
\item {\em Type-2 Critical Errors:} A correct \target $\ClassId$ is never matched, before or after the EOD.
\item {\em Type-3 Critical Errors:} An incorrect \target $\ClassId$ is matched before EOD. 
\end{packeditemize}
A useful property of this categorization is that any wrong decision (a correct \target not being picked, or an incorrect \target being picked) falls in a unique category, and non-critical errors do not belong to any category. 
In other words, as long as the loss function penalizes the occurrences of each category, it will only capture critical errors.

\mypara{\tool's first attempt of a new loss function}
To understand why it is difficult to penalize critical errors and critical errors {\em only}, we first consider the common practice of assigning a higher weight to the loss of a DNN output if the ground-truth of the input will lead to a selection of some \targets (\eg~\cite{mcdnn,jung2015joint,vasan2020imcfn}).
Henceforth, we refer to this basic design of loss function as \toolbasic.

At best, \toolbasic might improve the DNN's {\em label-wise} accuracy on inputs whose ground-truth decision selects some \targets.
However, as elaborated in \S\ref{subsec:study-mech}, we also need to consider which labels belong to the same \target, the decision type, and the matching order of an application decision process in order to capture the three types of critical errors.
For instance, in the garbage-classification application (Figure~\ref{fig:input_example}), without knowing the label lists of each \target, \toolbasic will give an equal penalty to a critical error of mis-classifying a \code{Paper} image to \code{Wood} and a non-critical error of mis-classifying a \code{Paper} image to \code{Shirt}.
Similarly, without knowing the matching order, \toolbasic will equally penalize the output of [\code{Plastic}, \code{Jacket}] and [\code{Jacket}, \code{Plastic}], but only the latter leads to correct output because \code{Jacket} is matched first.

\mypara{\tool's loss function} 
\tool leverages the categorization of critical errors to systematically derive a loss function that penalizes each type of critical error.
To make it concrete, we explain \tool's loss function of ``label-based API, \term{Multi-Choice} type of decision, and \term{App-order}'' (\eg Figure~\ref{fig:loss}).
Appendix\S\ref{app:loss_appendix} will detail the loss functions of other decision processes.
The loss function of such applications has three terms, each penalizing one type of critical error:

\vspace{-0.3cm}
{\small
\begin{align}
     L(\y) &=   
  \overbrace{ \textrm{Sigmoid}\left(\min\left(\max_{\LabelId \in \cup_{\ClassId < \hat{\ClassId}}\TargetSet_{\ClassId}}\y[\LabelId],\max_{\LabelId\in\TargetSet_{\hat{\ClassId}}}\y[\LabelId]\right)-\theta\right)}^{\substack{\text{\sf \footnotesize {{\bf Type-1} Critical Errors}} }} \label{eq:loss_app} \\&+
 \overbrace{ \textrm{Sigmoid}\left(\theta-\max_{\LabelId\in\TargetSet_{\hat{\ClassId}}}\y[\LabelId]\right)}^{\substack{\text{\sf \footnotesize {{\bf Type-2} Critical Errors}} }} + 
 \overbrace{ \sum_{\ClassId<\hat{\ClassId}}\textrm{Sigmoid}\left(\max_{\LabelId\in\TargetSet_{\ClassId}}\y[\LabelId]-\theta\right)}^{\substack{\text{\sf \footnotesize {{\bf Type-3} Critical Errors}} }}\nonumber
\end{align}
}
Here, $\y[\LabelId]$ denotes the score of the label $\LabelId$, $\TargetSet_\ClassId$ denotes the set of labels of \target $\ClassId$, $\hat{\ClassId}$ denotes the correct (\ie ground-truth) \target, and the sigmoid function $\textrm{Sigmoid}(x)=\frac{1}{1+e^{x}}$ will incur a higher penalty on a greater positive value.

{\em Why does it capture the critical errors?} 
Given this application is \term{Multi-Choice}, the EOD will occur right after the first match of a \target, \ie the first check with a $\ClassId$ such that $\max_{\LabelId \in \TargetSet_{\ClassId}}\y[\LabelId]\geq\theta$.
\begin{packeditemize}
\item A Type-1 critical error occurs, if (1) the correct \target $\hat{\ClassId}$ is matched {\em and} (2) it is matched after the EOD.
First, the correct \target $\hat{\ClassId}$ is matched, if and only if at least one of its labels has a score above the confidence threshold, so $\max_{\LabelId\in\TargetSet_{\hat{\ClassId}}}\y[\LabelId]\geq\theta$).
Second, this match happens after the break, if and only if some \target $\ClassId$ before $\hat{\ClassId}$ (\ie $\ClassId<\hat{\ClassId}$) is matched, so $\max_{\LabelId \in \TargetSet_{\ClassId}}\y[\LabelId]\geq\theta$).
Put together, the first term of Eq~\ref{eq:loss_app} penalizes any occurrence of these conditions. 
\item A Type-2 critical error occurs, if no label in the correct \target $\hat{\ClassId}$ has a score high enough for $\hat{\ClassId}$ to be matched, \ie $\max_{\LabelId\in\TargetSet_{\hat{\ClassId}}}\y[\LabelId] < \theta$, so the second term of Eq~\ref{eq:loss_app} penalizes any occurrence of this condition.
\item A Type-3 critical error occurs, if any incorrect \target $\ClassId$ before $\hat{\ClassId}$ (\ie $\ClassId<\hat{\ClassId}$) has a label with a score high enough for $\ClassId$ to be matched, \ie $\max_{\LabelId\in\TargetSet_{\ClassId}}\y[\LabelId]-\theta$, so the third term of Eq~\ref{eq:loss_app} penalizes any occurrence of this condition.
\end{packeditemize}

To train a DNN, the loss function must be differentiable with respect to the DNN ouput $\y$. 
Eq~\ref{eq:loss_app} uses the $\max$ function several times. 
Though $\max$ is not naturally differentiable, it can be closely approximated in well-known differentiable forms provided by PyTorch's differentiable operators~\cite{paszke2017automatic}).

\tightsubsection{Extracting applications' decision process}
\label{sec:extract}

The current prototype of \tool program analysis supports Python applications that make decisions based on categorical label output or floating point output of ML APIs. We first discuss how it works for ML APIs with categorical label output, 
like all the APIs in Table \ref{tab:all_stats} except for \code{analyze\_sentiment}. We will then discuss a variant of it that works for most use cases of \code{analyze\_sentiment}.

Given application source code, \tool first identifies all the invocations of ML APIs. For every invocation $I$ in a function $f$, \tool then identifies all the branches whose conditions have a data dependency upon the ML API's label output. We will refer to these branches as $I$-branches. If there is no such branch in $f$, \tool then checks the call graph, and analyzes up to 2 levels of callers and up to 5 levels of callees of $f$ until such a branch is identified. If no such branch is identified after this, \tool considers the ML API invocation $I$ to not affect application decisions and hence does not consider any optimization for it. If some $I$-branches are identified, \tool records the top-level function analyzed, $F$, and moves on to extract the \summary in following steps.

\myparashort{What are the \targets?} 
\tool figures out all the \targets and their composition in two steps.

The first step leverages symbolic execution and constraint solving to identify all the labels that belong to \textit{any} \targets. Specifically, \tool 
applies symbolic execution to function $F$, treating the parameters of $F$ and the
label output of $I$ as symbolic (\ie the symbolic execution skips the ML API invocation $I$ and directly uses $I$'s symbolic output in the remaining execution of $F$)\footnote{{Recall that an API output contains several fields not used to influence 
control flow in any applications. We set them with pre-defined dummy values.}}. 
{
Since applications typically match only one label in API output at a time (as observed in \S\ref{subsec:study-mech}), we set the label array returned by $I$ to contain one element (label) and use a symbolic string to represent it. 
}
Through symbolic execution, \tool obtains constraints for every path that involves an $I$-branch, solving which tells \tool which labels need to be in the 
output of the ML API in order to execute each unique path, essentially all the labels that belong to any \target.

One potential concern is that a solver may only output one instead of all values that satisfy a constraint.
Fortunately, the symbolic execution engine used by \tool, NICE~\cite{irlbeck2015deconstructing}, turns Python code into an intermediate representation where each branch is in a simplest form. Take Figure \ref{fig:code}(d) as an example, the source-code branch 
\code{if obj.name in Protein} is transformed into three branches where \code{obj.name}
is compared with \code{``Hamburger''}, \code{``Meat''}, and \code{``Patty''} separately, allowing us to capture all three labels by solving three separate path constraints.

The second step groups these labels into \targets by comparing their respective
paths: 
{if two API output, each with one label, lead}
the program to follow the same execution path at the source-code level, these two labels belong to the same \target.
For example, in Figure \ref{fig:code}(d), the execution path is exactly the same when the \code{label\_detection}
API returns \code{[``Hamburger'']}, comparing with when it returns \code{[``Meat'']}, with
all function parameters and other API output fields being the same. Consequently, we can know that 
label \code{Hamburger} and label \code{Meat} belong to the same \target.
To figure out the path, \tool simply executes function $F$ using each input produced by the constraint solver and traces the source-code execution path using the Python trace module.

One final challenge is that \tool needs to identify and exclude the path where none of the \targets are matched (\eg the \code{``It is others.''} path in Figure \ref{fig:code}(a)).
We achieve this by carefully setting the default solution in the constraint solver to be an empty string, 
which
is impossible to output for any ML APIs in this paper. This way, whenever this default solution is output, \tool knows that the corresponding path 
matches no \target.

\myparashort{What is the type of decision?}
When only one \target is identified, \tool reports a \term{True-False} decision type. Otherwise, \tool decides whether the decision type is 
\term{Multi-Choice} or \term{Multi-Select} by checking the source-code execution path associated with every \target label obtained above. 
If any execution evaluates an $I$-branch \textit{after} another $I$-branch
is already evaluated to be true, \tool reports a 
\term{Multi-Select} decision type; otherwise, \tool reports a \term{Multi-Choice}
decision type.

\myparashort{What is the matching order over the \targets?}
To tell whether a \term{Multi-Choice} decision is made
through \term{API-Order} like in Figure \ref{fig:input_example} or 
\term{App-Order} like in Figure \ref{fig:code}(a), \tool first identifies all the \code{for} loops that
iterate through the label array output by the ML API and have control-dependency
with $I$-branches, \eg 
the \code{for l in labels} in Figure~\ref{fig:code}(a)
and the \code{for obj in response.label\_annotations} in Figure~\ref{fig:input_example}.

\tool then checks how many such output-iterating loops there are.
If there is only one and this loop is not inside another loop, 
like that in Figure \ref{fig:input_example}, \tool considers the matching order to be
\term{API-Order}, as the application only iterates through
each output label once, with the matching order determined
by the output array arranged by the ML API. Otherwise, \tool considers
the matching order to be \term{App-Order}. This is the case for the example
shown in Figure \ref{fig:code}(a), where three output-iterating loops are
identified, each of which matches with one \target in an order determined by the
application: the \code{Landmark} \target, followed by the 
\code{Building}, and
finally the \code{Person}.

\myparashort{How to handle floating-point output of ML APIs?}
Recall in \S\ref{subsec:study-mech} that some ML APIs, \eg \code{analyze\_sentiment}, have floating-point output and the application defines several value ranges to put each floating-point output into one category. To handle this type of API, \tool needs to identify the value
range of each \target, which is not supported by NICE and other popular constraint solvers. Fortunately, many applications directly compare API output with constant values in $I$-branches, giving \tool a chance to infer the
value range. For these applications, \tool first extracts those constant values that are compared with API output in $I$-branches, 
\eg 0.3 and 0.6 in Figure \ref{fig:code}(b).
\tool then forms tentative value ranges using these numbers, like -1 -- 0.3, 0.3 -- 0.6, and 0.6 -- 1 for Figure \ref{fig:code}(b) (-1 and 1 are the smallest and biggest possible \code{score} output of \code{analyze\_sentiment} based on the API manual). To confirm these value ranges and figure out the boundary situation, \tool then
executes function $F$ with all the boundary values, as well as some values in the middle of each range. By comparing which values lead to the same execution path, \tool finalizes the value ranges. For the example in Figure \ref{fig:code}(b), after executing with \code{score} set to -0.35, 0.3, 0.45, 0.6, and 0.8, \tool settles down on the final value ranges to be: (-1,0.3), [0.3,0.6), and [0.6,1).

\myparashort{Limitation}
The static analysis in \tool does not handle the iterated object of while loops, unfolded loops, and recursive functions. For complexity concerns, \tool only checks caller and
callee functions with limited levels, and hence may miss some $I$-branches
far away from the API invocation. \tool's ability of identifying \targets is limited by the constraint solver. \tool assumes different source-code paths correspond to different target classes, which in theory could be wrong 
if the application behaves exactly the same under different execution paths.

\tightsubsection{Putting them together}
\label{sec:runtime}

\begin{figure}
     \includegraphics[width=.99\linewidth]{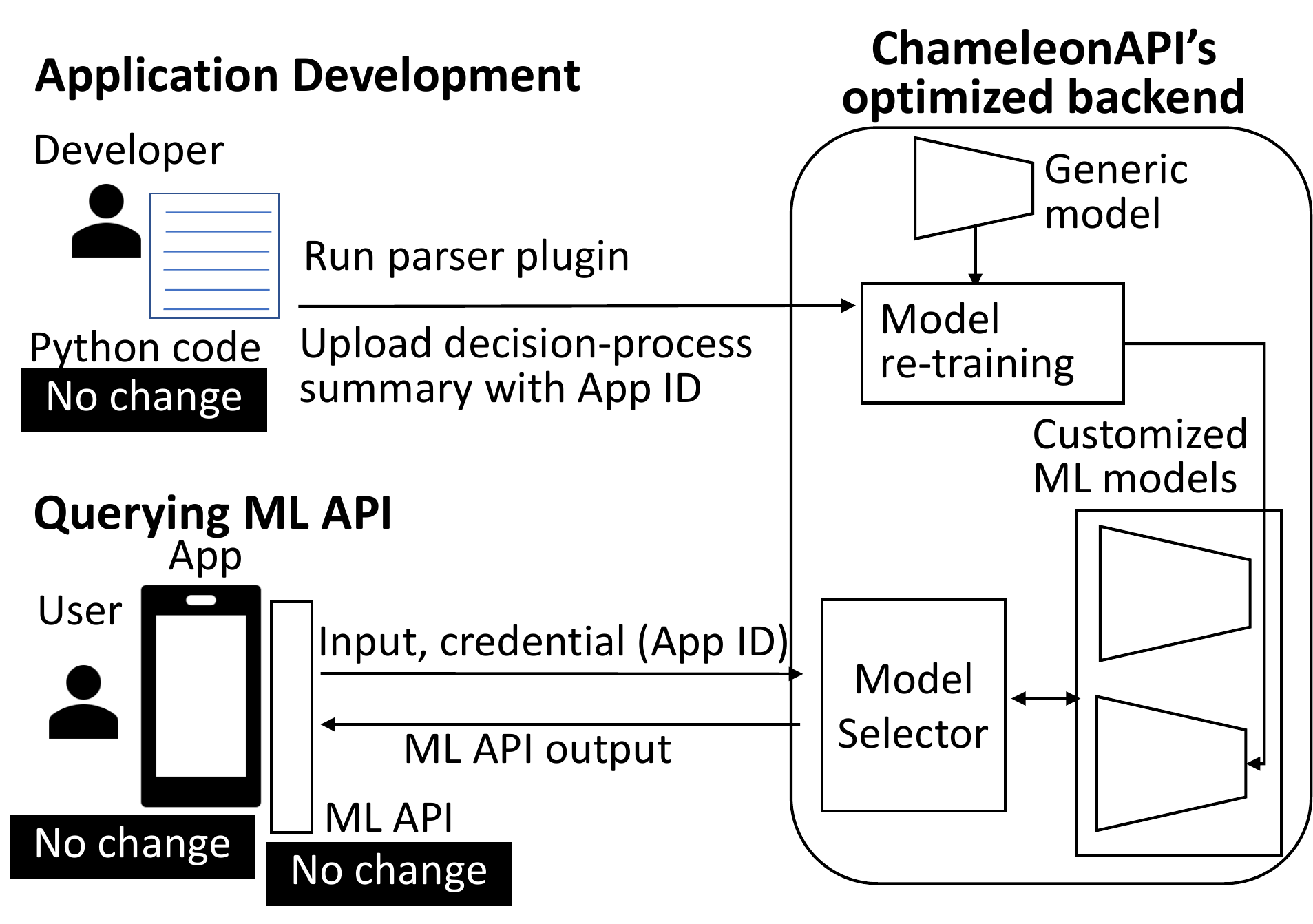}
    \tightcaption{Workflow of \tool.
    }
\label{fig:system}
\end{figure}

We put these components together into a ML-as-a-Service workflow shown in Figure~\ref{fig:system}. 

First, when an application (\App) is developed \edit{or updated}, the developers run a parser (described in \S\ref{sec:extract}) provided by \tool on \App's source code to extract the \summary for \App.
The developers can then upload the \summary to \tool's backend together with a unique application ID\footnote{In many MLaaS offerings~\cite{google-cloud,amazon-ai}, a connection between the application and the MLaaS backend is commonly created before the application issues any queries. Existing MLaaS already allows applications to specify the application ID via the connection between the application and backend.} (which will later be used to identify queries from the same application). 

\tool's backend then uses the received \summary to construct a new  application-specific loss function (described in \S\ref{sec:loss}). 
When a DNN is trained using the new loss function, its inference results will lead to fewer critical errors (\ie incorrect application decisions) for application \App.
In our prototype, \tool uses the new loss function to re-train an off-the-shelf pre-trained DNN, a common practice to save training time (see \S\ref{sec:evaluation} for quantification).
The DNN re-training uses 
an application-specific dataset sampled from the dataset \edit{used by the pre-trained
 generic DNN} (see Table~\ref{tab:datasets} and \S\ref{sec:implement}), so that each target/non-target class is selected by ground-truth decisions of the same number of inputs.

Finally, \edit{\tool backend maintains a set of DNN models, each customized for an application and keyed by the application ID. When application \App invokes an ML API at run time, the \tool backend
will use the application ID associated with the API query to identify the DNN model customized for \App, run the DNN on the input, and return the inference result of the selected model to the application.
}

\edit{Note that, \tool can also be used to customize ML models that run locally behind the ML APIs, instead of those in the
cloud through ML service providers. In this case, developers run the \tool parser on their application
and save the parser's result into
a local file. This local file will then be consumed to help re-train an off-the-shelf DNN into a
customized DNN to serve the application.}



%% file: 4-implement.tex
\label{sec:implement}

\mypara{Extractor of \summary}
The current prototype of \tool is implemented for Python applications that use Google or Amazon ML APIs.
It takes as input the application source code and returns as output the \summary in the JSON format.
It uses NICE symbolic execution engine~\cite{irlbeck2015deconstructing} and CVC5 constraint solver~\cite{cvc5} to identify \targets, and uses Python static analysis framework Pyan \cite{pyan} and Jedi \cite{jedi} to identify the decision type and the matching order. 
Particularly, it identifies the object that is iterated through by a \code{for}-loop through the \code{iter} expression in each for-loop header, which is used to distinguish \term{Multi-Choice} and \term{Multi-Select} decisions and the matching order.

\mypara{ML re-training} 
The re-training module is implemented in PyTorch v1.10 and CUDA 11.1.
It uses a \summary to construct a new loss function (see \S\ref{sec:loss}), and then 
replaces the builtin loss function in Pytorch with the new loss function, and uses the common forward and backward propagation procedure to re-train an off-the-shelf pre-trained DNN model (explained next).

\begin{table}[]
\centering
\begin{footnotesize}

\begin{tabular}{llll}
\hline
                         & Dataset       & \Genmodel \\ \hline
{\begin{tabular}[l]{@{}l@{}}Image Classification\end{tabular}}       &  OpenImages~\cite{openimages}    & TResNet-L~\cite{ASL}        \\ \hline
{\begin{tabular}[l]{@{}l@{}}Object Detection\end{tabular}}     &  COCO~\cite{coco}          & Faster-RCNN~\cite{NIPS2015_14bfa6bb}      \\ \hline
{\begin{tabular}[l]{@{}l@{}}Sentiment Analysis\end{tabular}}   &  Amazon review~\cite{marc_reviews} & BERT~\cite{devlin2018bert}              \\ \hline
{\begin{tabular}[l]{@{}l@{}}Text Classification\end{tabular}}  &  Yahoo~\cite{yahoo}  & BERT~\cite{devlin2018bert}             \\ \hline
{\begin{tabular}[l]{@{}l@{}}Entity Recognition\end{tabular}}   & conll2003 ~\cite{tjong-kim-sang-de-meulder-2003-introduction}      & BERT~\cite{devlin2018bert}               \\ \hline
\end{tabular}
\tightcaption{The ML APIs and datasets  in evaluation. }
\label{tab:datasets}
\end{footnotesize}
\end{table}

\mypara{Generic models} 
Without access to \edit{the models and the training data used by commercial ML services}, we use open-sourced pre-trained DNNs and their training datasets as a proxy, which are summarized in Table~\ref{tab:datasets}. These DNNs are trained on the ``training'' portion of their respective datasets. They are trained to achieve good accuracy over a wide range of labels, and we have confirmed that their accuracies in terms of application decisions are similar to the real ML APIs (\S\ref{subsec:results}). 

\mypara{Training data}
We make sure that the labels included in these datasets cover the labels used in the decision processes of the applications in our study.
An exception is text classification: to our best knowledge, there is no open-source dataset that covers the classes in Google's text classification API. 
Instead, we use the Yahoo Question topic classification dataset~\cite{yahoo}, whose classes are similar to those used in the applications.

Instead of training DNNs on all training data, most of which do not match any \targets of an application, we create a downsampled training set for \tool and \toolbasic.
For each application, we randomly sample (without replacement) its training data such that each \target and the non-target class (not matching any \target) is the correct decision for the same number of training inputs, which depending on applications, ranges from 12K to 40K.
With such training set, \toolbasic 
will be equivalently implemented by training on the downsampled training set using the conventional loss function (\ie cross-entropy loss for classification tasks). 
Moreover, the downsampled training set significantly speedups DNN re-training (\S\ref{subsec:results}).

%% file: 5-evaluation.tex
\label{sec:evaluation}
Our evaluation aims to answer following questions:
How much can \tool reduce incorrect application decisions? 
How long does it take \tool to customize DNN models for applications? 
and Why does \tool reduce incorrect application decisions where \toolbasic falls short?

\tightsubsection{Setup}
\label{sec:eval_setup}

\begin{figure*}
\centering
    \subfloat[][Applications making \term{True-False} decisions.]
    {
        \includegraphics[width=0.99\textwidth]{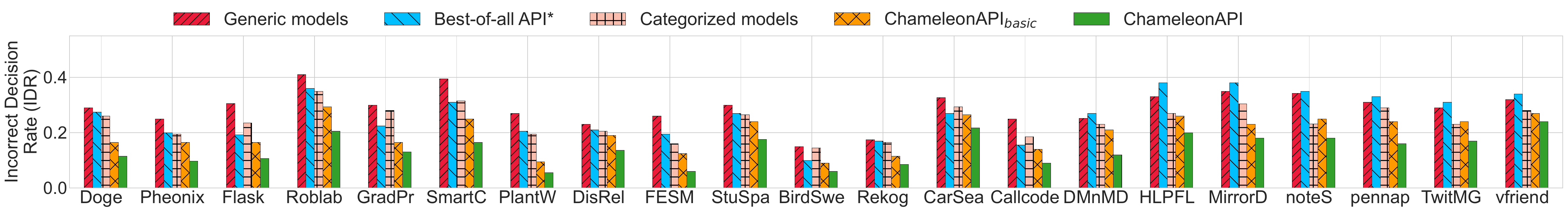} 
        \label{fig:tf}
    }
    
    \subfloat[][Applications making \term{Multi-Choice} decisions.]
    {
        \includegraphics[width=0.95\textwidth]{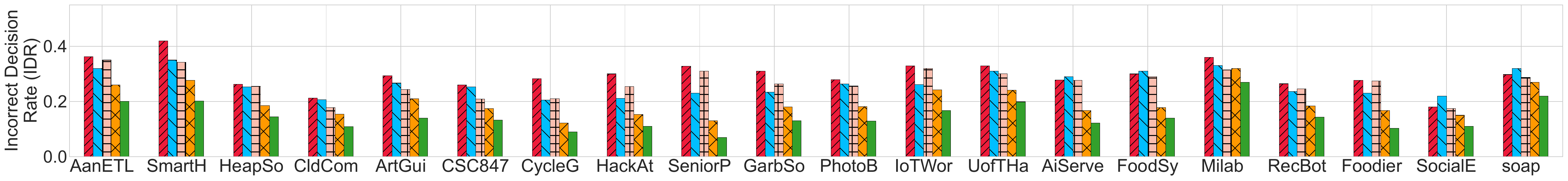} 
        \label{fig:multi_choice}
    }

    \subfloat[][Applications making \term{Multi-Select} decisions.]
    {
        \includegraphics[width=0.92\textwidth]{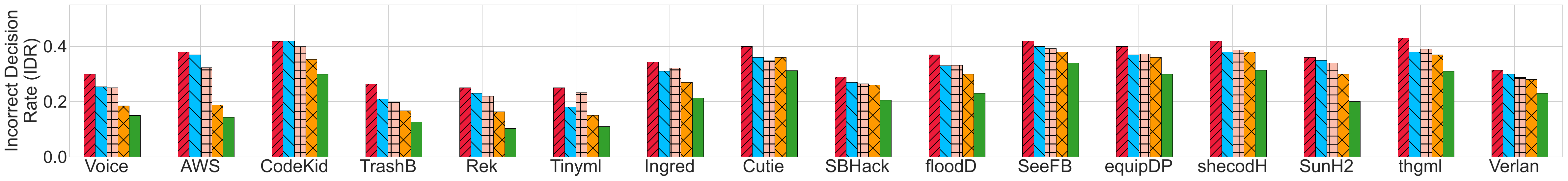} 
        \label{fig:multi_select}
    }
    \tightcaption{\edit{\tool reduces the incorrect decision rate (\Rate) on the 57 applications that use Google's or Amazon's image-classification, text-classification, and object-detection APIs. 
    } 
    }
    \label{fig:overall}
\end{figure*}

\mypara{Applications} 
\edit{We have applied \tool on all the 77 applications summarized in Table \ref{tab:all_stats}.
Due to space constraints, our discussion below focuses on the 57 applications 
 that involve three popular ML tasks, image-classification, object-detection, and text-classification, and omits the remaining 20 
 applications that involve sentiment analysis and entity
 recognition. The results of the latter show similar trends of advantage from \tool and are available in Appendix \S\ref{appendix:appendix_otherapp}. 
 }

\mypara{Metrics} 
For each scheme (explained shortly) and each application, we calculate the \emph{incorrect decision rate} ({\em \Rate}): the fraction of testing inputs whose application decisions do not match the correct application decisions (\ie decisions based on human-annotated ground truth).

\mypara{Schemes} We compare the results of these schemes: 
\begin{packeditemize}
    \item \emph{Various commercial ML APIs}: the results returned by ML APIs of three \edit{service} providers (Google~\cite{google-cloud}, Amazon~\cite{amazon-ai} and Microsoft~\cite{ms-azure}). 

    \item \emph{\bestapi}: a hypothetical method that queries ML APIs from those three service providers on each input and picks the best output based on the classic definitions of accuracy: label-wise recall for classification tasks and mean-square-error of floating-point output for sentiment analysis. This serves as an idealized reference of recent work~\cite{chen2020frugalml,chen2022frugalmct}, which tries to select the best API output with high label-wise accuracy.
    \item \edit{\emph{\Genmodels}: the open-sourced \genmodel} based on which \edit{the next three schemes} are re-trained. They serve as a reference without customization and achieve similar accuracy as commercial APIs. \edit{Their details are explained in Section \ref{sec:implement}.}
    \item \edit{\emph{Categorized models}: 
This scheme pre-trains a number of specialized models.
Each specialized model replaces the last layer of the generic model 
so that it outputs the confidence scores for a smaller number of labels representing a common category
\jc{is this a typo? we are not replacing the last layer with some labels, right?}(e.g., ``dog'', ``animal'', ``person'' and a few other labels represent the ``natural object'' category), and is
fine tuned from the generic model accordingly. 
A simple parser checks which labels are used by an application. 
If all the labels belong to one category, the corresponding model
specialized for this category is used to serve API calls from this application.
If the labels belong to multiple categories, multiple specialized models will be
used, which we will explain more later.
\jc{can you give a bit more detail on how this mapping is done?} 
We set up 35 categories for image classification
and 7 categories for object detection based on the
Wikidata knowledge graph \cite{wikidata}, as well as 15 categories for text classification
based on the inherent hierarchy in Google text-classification output. More details of how we have designed these categories are available in
the Appendix \S\ref{appendix:appendix_complete}.

    Note that, we have designed this scheme to represent a middle-point in the design space
    between the generic model and the \tool approach: 
    on one hand,
    this scheme offers some application
    customization, but not as much as \tool (e.g., which labels belong to the 
    same \target, what is the decision process, and what is the matching order used
    by the application are
    all ignored);
    \jc{maybe say explicitly what's customized in \tool but not in this approach?}
    on the other hand, this scheme requires a simpler parser compared to \tool.
     }

    \edit{
    \item \emph{\modeld}: the model is re-trained with \tool's training data, which concentrates on labels used by the application, but with the conventional loss function. Like \prespec, this scheme only needs a simple parser
    that extracts which labels are used by the application, and does not make use of
    other application information that \tool uses. Unlike \prespec, this scheme
    prepares a customized model for each application, instead of relying on a
    small number of categorized models.
    \jc{maybe its a good idea to clarify the diff between categorized model vs. \modeld}
    }
    \item \emph{\tool} (our solution): the model re-trained with our training data and loss function.
\end{packeditemize}

\begin{figure}
    \centering
    \includegraphics[width=0.44\textwidth]{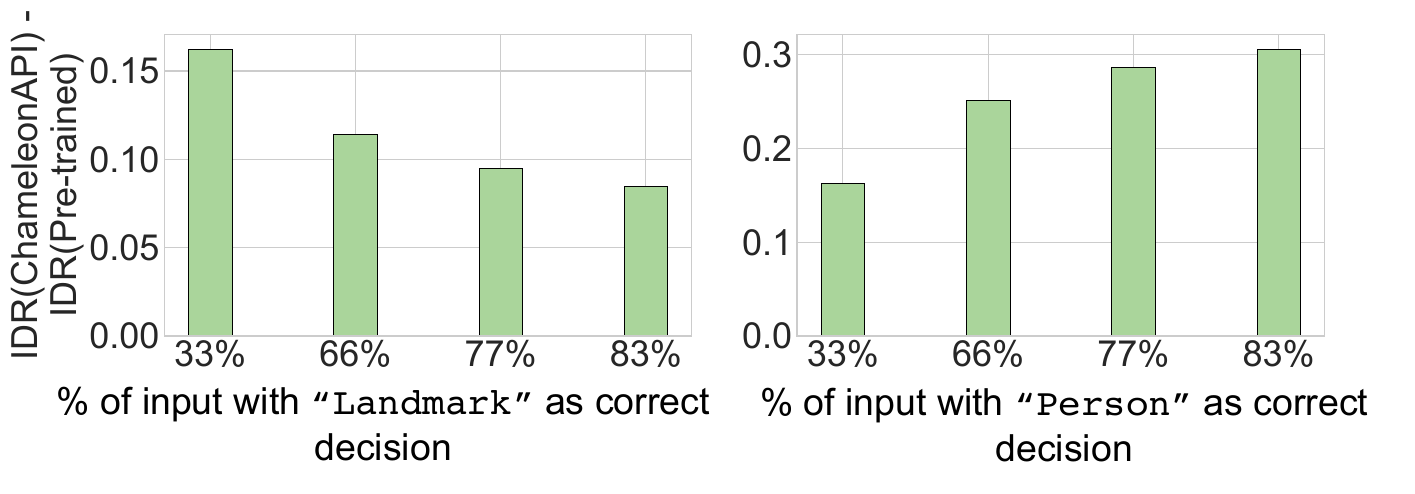} 
    \tightcaption{How the accuracy advantage of \tool changes with different input distribution.
    }
    \label{fig:ood}
\end{figure}

\mypara{Testing data} 
For the same application, all schemes are tested against the same testing input set. The testing set of an application is randomly sampled from the ``testing'' portion of the dataset associated with the application's \edit{\genmodel} (Table~\ref{tab:datasets}). We make sure that {\em no} testing input appears in the training data. Like the creation of training data of \tool (\S\ref{sec:implement}), by default, we randomly sample the testing data such that each \target and the non-target class (not matching any \target) appear as the correct decision for the same number of testing inputs, which ranges from 1.2K to 4K. This is similar to the testing sets used in related work on ML API (\eg~\cite{kang2018blazeit,kang2017noscope,wan2022automated, chen2022frugalmct}). Such data downsampling is commonly used in ML~\cite{elrafey2017recent,liu2020study}. Other than Figure~\ref{fig:ood}, we will use this as the default testing dataset.

\mypara{Hardware setting}
We evaluate \tool and other approaches on a GeForce RTX 3080 GPU, and an Intel(R) Xeon(R) E5-2667 v4 CPU, with 62GB memory.

\begin{table}[]
\setlength\tabcolsep{5pt}
\centering
\begin{tabular}{@{}rccc@{}}
\toprule
                  & \term{True-False} & \term{Multi-Choice} & \term{Multi-Select} \\ \midrule
Google API        & 0.29       & 0.32         & 0.35         \\
Microsoft API     & 0.30       & 0.33         & 0.32         \\
Amazon API        & 0.31       & 0.33         & 0.36         \\
\bestapi         & 0.26       & 0.27        & 0.31         \\
\hline
\Genmodels & 0.29       & 0.30         & 0.34         \\
\prespec & 0.24 & 0.27 & 0.31 \\
\toolbasic  & 0.19       & 0.22         &   0.27        \\
{\bf \tool}              & {\bf 0.13}       & {\bf 0.16}        & {\bf 0.21}         \\ \bottomrule
\end{tabular}
\tightcaption{\edit{Average incorrect decision rate (\Rate) among apps. that make different types of decisions.} The lower the better. The top half represents commercial APIs and their idealistic combinations; the bottom half represents open-source models.
}
\label{tab:overall}
\end{table}

\begin{table}[]
\setlength\tabcolsep{5pt}
\centering
\begin{tabular}{rccc}
\hline
                   & Single-category  & Multi-category  \\ \hline
\Genmodels & 0.32                 & 0.28               \\
\prespec & 0.28                 & 0.27               \\
\toolbasic & 0.24 & 0.18 \\
ChameleonAPI       & 0.17                 & 0.14                \\ \hline
\end{tabular}
\tightcaption{\edit{Average \Rate among single-category and multi-category applications. The lower the better.}}
\label{tab:break_category}
\end{table}

\tightsubsection{Results}
\label{subsec:results}

\mypara{Overall gains}
\edit{
Measured by the average incorrect decision rate (IDR) across all applications, the most
accurate scheme is \tool, with an IDR of 0.16, 
and the least accurate scheme is \Genmodels, with an IDR of 0.31. In other words, \tool 
successfully reduces the number of incorrect decisions of its baseline model by almost 50\%.
\toolbasic (0.22), \prespec (0.28), and \bestapi (0.28) have IDR rates in between. 

The advantage of \tool, and even \toolbasic, over the other schemes is consistent across
all three types of applications that make different types of decisions, as shown in 
Table \ref{tab:overall}. In fact, \tool offers the highest accuracy by a clear margin
for every single application in our evaluation, as shown in Figure \ref{fig:overall}.

To better compare the \tool approach with \prespec,
we divide the 57 applications into two
types: (1) 39 single-category applications --- each
application uses labels that belong to one category and hence can benefit from one specialized model in the \prespec scheme; (2)
18 multi-category applications --- each application uses labels that
belong to multiple categories. For these applications, the \prespec scheme feeds the API input to multiple specialized models and combines
these models' output to form the API output. 
As shown in Table \ref{tab:break_category}, the \prespec scheme does offer improvement
from \Genmodels by considering which labels belong to an application's \targets, particularly for single-category applications. However, both \tool and \toolbasic
perform better than \prespec for both single-category and
multi-category applications---the per-application customization in \tool
and \toolbasic has paid off. 

The above advantage of \tool over \toolbasic and \prespec 
shows that the static analysis used in \tool to
extract not only what labels are used by the application, but also which
labels belong to the same \target, the decision type, and the matching order, as described in Section \ref{sec:extract}, 
is worthwhile.\jc{this sentence sounds repetitive to the end of the last para} \shan{I have edited the last sentence of the previous paragraph. Does it
look OK now?}
}

\begin{figure}
    \centering
     \includegraphics[width=.86\linewidth]{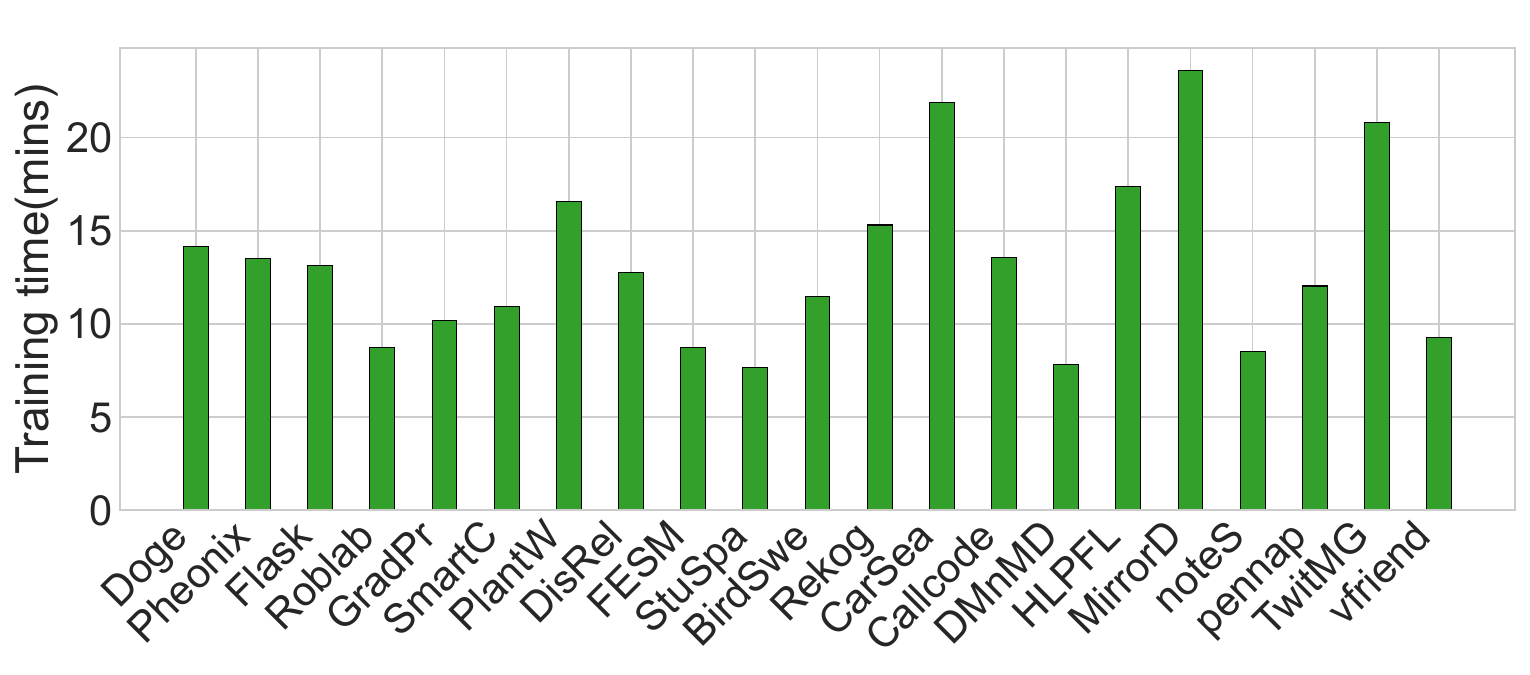}
    \tightcaption{ Re-training time for applications in Figure~\ref{fig:overall}(a). 
    }
\label{fig:google_time}
\end{figure}

\edit{
\mypara{Cost of obtaining customized models}
The customization effort of \tool includes two parts (1) extracting the \summary from application source code, and (2) re-training the ML model. The first part takes a few seconds: on an Intel(R) Xeon(R) E5-2667 v4 CPU machine, our parser extracts the \summary from every benchmark application within 10 seconds.

The second part takes a few minutes, much faster than training a neural network from scratch. As shown in
Figure~\ref{fig:google_time}, re-training DNNs for the 21 applications in Figure~\ref{fig:overall}(a) on a single RTX 3080 GPU takes 8 to 24 minutes.
Focusing on a small portion of all possible labels (\S\ref{subsec:study-imp}),
\tool fine-tunes pre-trained models using much less training data
 than the \edit{\genmodels} 
and thus needs fewer iterations to converge. 

Considering that a V100 GPU with similar processing GFLOPS as our RTX 3080 GPU only costs \$2.38 per hour on Google Cloud \cite{google-price}, re-training an ML model for one application costs less than \$1.} 

\edit{

\mypara{Cost of hosting customized models} For cloud providers,
\tool would incur a higher hosting cost than traditional
ML APIs by serving a customized DNN for every application instead of
a generic DNN for all applications.

The extra cost includes more disk space to store customized neural
network models. 
For example, each image-classification model in \tool uses 115 MB of disk space. So, for $n$ applications, $115 \cdot n$ MB of disk space 
may be needed to store \tool customized models.

The extra cost also involves more GPU resources. A naive design of 
using one GPU to exclusively serve requests to one customized neural network model
will likely lead to underutilization of GPU resources. To serve 
different applications' customized models on one GPU, we need to pay
attention to memory working set and performance isolation issues.
In our experiments on an RTX 3080 GPU, 
loading an image-classification model from CPU to GPU RAM takes 18 to 40
ms (inference itself takes 10 to 35 ms with a batch size of 1).
Fortunately, modern GPU has sufficiently
large RAMs to host several requests to different customized models 
simultaneously: in our experiments, the peak memory consumption of 
one inference request is less than 2GB. Furthermore, the majority of
the model inference memory consumption comes from intermediate states, instead of the model itself. Consequently, the memory consumption of
multiple inference requests on different models is similar to that on the same model.


Of course, \tool can take advantage of recent proposals to improve GPU sharing~\cite{shepherd, NVIDIAMPS, salus, gemel} as well as to reduce the footprint of serving multiple DNNs~\cite{mainstream}.
These techniques could be advantageously employed by \tool to determine the optimal degree of sharing among customized DNNs, and we leave them
to future work.

Finally, there is also the extra cost of needing more complex software to manage the DNN serving.
For instance, \tool needs to dynamically route each request to a GPU that serves the DNN of the application (see \S\ref{sec:runtime}). 



}


\begin{figure}
    \centering
     \includegraphics[width=.99\linewidth]{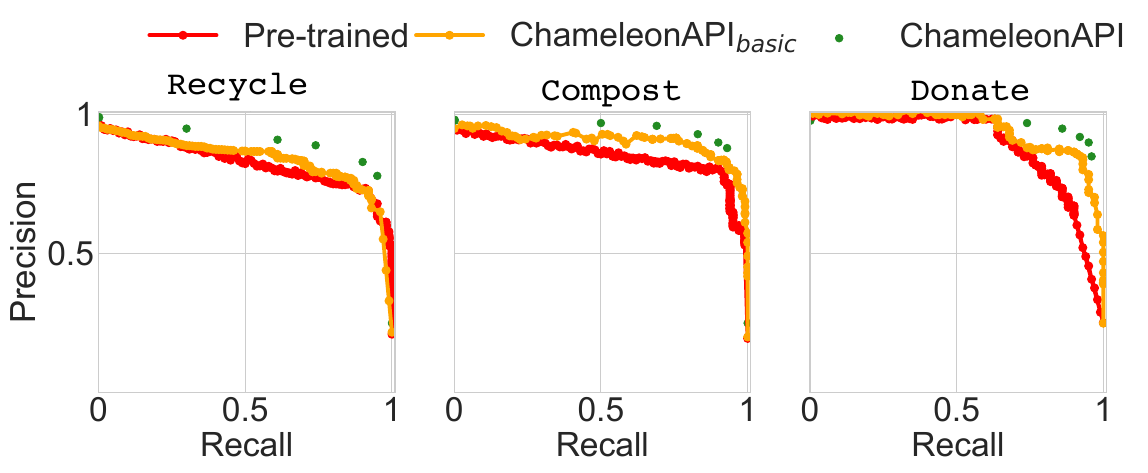}
    \tightcaption{ Precision-recall trade-off for \code{HeapsortCypher}. }
\label{fig:pr}
\end{figure}

\mypara{Precision-recall tradeoffs}
Traditionally, for a trained ML model, it is common to vary the confidence-score thresholds in order to find the best precision-recall tradeoff of a trained model. Thus, it is important that \tool also achieves better precision-recall tradeoffs. Figure~\ref{fig:pr} shows the precision-recall results in each  \target of a particular application, by varying the detection threshold $\theta$ (defined in \S\ref{sec:loss}) of two baselines (real APIs are excluded, because we cannot change their thresholds and their \Rate is not as low as \toolbasic).  \tool's tradeoffs are better than both baselines (and we observe similar results in other applications). Note that since \tool's loss function uses an assumed $\theta$, we do not vary the $\theta$ when testing it; instead, we re-train five DNNs of \tool, each with a different $\theta$ and test them with their respective thresholds.

\mypara{Understanding the improvement}
\tool's unique advantage is that it factors in the decision process of an application, including not only the \targets but also the decision type and the matching order.
Next, we use two case studies to further reveal the underlying tradeoffs made by \tool
to achieve its improvement on application-decision accuracy.

First, \edit{
\tool reduces errors related to different \targets differently depending
on their different roles in the application decision process}. This effect is particularly striking in \term{Multi-Choice} applications with the matching order of \term{App-Order}, where the first \target is always matched against API output. Thus, when the correct \target is not the first one, falsely including a label that belongs to the first \target will more likely be a critical error than other mis-classifications, because it will block the match of other \targets. To illustrate this, we consider the \term{Multi-Choice} application of \code{Aander-ETL}. We increase the percentage of testing inputs whose correct action is the first \target or the last \target. Figure~\ref{fig:ood} shows that increasing the portion of inputs of the last \target (\code{Person}) generally leads to bigger gains of \tool, whereas increasing the portion of the first \target (\code{Landmark}) does the opposite. \edit{This shows the application itself
already has good tolerance to mis-classification on inputs that belong to the first \target, but not to mis-classification on the inputs that belong to later
\targets, which is exactly where \tool can help.}

Second, recall from \S\ref{sec:loss} that our loss function helps to minimize critical errors, even at the cost of missing labels that do not affect application decisions (\ie non-critical errors). To show this, we define {\em label error rate} on an image as the fraction of the image's ground-truth labels that are missed by the DNN output (a label list). We consider \code{IoTWor} (explained in Table~\ref{tab:stats}), which similar to \code{Anander-ETL} makes \term{Multi-Choice} decisions with \term{App-Order} matching order. The average label missing rate of \tool on our testing images is 0.21, which is slightly higher than \toolbasic's 0.18. This means \tool makes more label-level mistakes than \toolbasic. However, our \Rate (0.17) is 44\% lower than \toolbasic, which means \tool makes far fewer critical errors.

%% file: 7-related_works.tex
\label{sec:related}

Due to space constraints, we discuss related papers that have not been discussed earlier in the paper.

\mypara{Optimizing storage and throughput of DNN serving}
Various techniques have been proposed to optimize the delay, throughput, and storage of ML models via model distillation~\cite{mullapudi2019online,shen2017fast,poe}, pruning~\cite{mcdnn} or cascading~\cite{cao2021thia,anderson2019physical}. 
This line of work explores a different design space than \tool: they design ML models with higher inference speed or smaller model size with minimum loss in accuracy. 
\tool focuses on re-training existing ML models such that the rate of incorrect decisions of a given application is reduced.

\mypara{Application-side optimization} Recent work also proposes to change the applications to better leverage existing \edit{ML APIs}. One line of work~\cite{chen2020frugalml,chen2022frugalmct,xie2022cost} invokes ML APIs from different service providers to achieve high accuracy within a query cost budget. 
Another line of work aims to eliminate  misuse of ML APIs in applications~\cite{wan2021machine,wan2022automated}. They require changes to the application source code (\eg changing the API input preparation, switching from image-classification API to object-detectin API, etc.). They are complementary to our work, because we customize the \edit{ML-API backend DNN} and do not require changes on the application's source code.

\mypara{Measurement work on MLaaS} For their rising popularity, ML-as-a-Service platforms have also attracted many measurement studies to understand accuracy~\cite{chen2021did}, performance~\cite{yao2017complexity}, robustness~\cite{hosseini2017google}, and fairness~\cite{koenecke2020racial}. However, they have so far not taken in account the ML applications that use ML APIs, and is thus different from our empirical study of ML applications in \S\ref{sec:empirical}. Previous work that studies ML applications \cite{wan2021machine} did not look into the decision making process and how ML API errors might affect different applications differently.

Finally, a myriad of techniques have been studied to better manage and schedule GPU resources in ML training/serving systems (\eg~\cite{clipper,infaas,inferline,mlaasinthewild,zhang2018deepcpu,grnn,zhang2019mark,parity,shen2019nexus,gujarati2020serving,lee2018pretzel,mohan2022looking,8980322,han2022microsecond,choi2022serving}).
They aim for different goals than \tool, but these techniques can be used to help \tool train and serve the application-specific ML models.

%% file: 8-conclusion.tex
\label{sec:conclusion}


\edit{ML APIs} are popular for its accessibility to application developers who do not have the expertise to design and train their own ML models. In this paper, we study how the \edit{generic ML models behind ML APIs} might affect different applications' control-flow decisions in different ways, and how some ML API output errors may or may not be critical due to the application decision making logic. Guided by this study, we have designed \tool that offers both the accuracy advantage of a custom ML model and the accessibility of the traditional \edit{ML API}.

%% file: 9-appendix.tex
\begin{appendices}

\onecolumn 
\section{Applications}
\label{app:appendix}

\newcolumntype{L}{>{\small\arraybackslash}l}
\newcolumntype{C}{>{\small\arraybackslash}c}
\setlength\tabcolsep{1.5pt}
\begin{tabularx}{\linewidth}{|L|C|C|L|}
\caption{The statistics of 77 applications in empirical study. \footnotesize{(Multi-Choice-* refer to Multi-Choice (*-Order).)}} \label{tab:stats}\\ 
\hline
\multicolumn{1}{|C|}{\begin{tabular}[c]{@{}c@{}}\textbf{Application name}\\{(Link to Github repo)}\end{tabular}} & \multicolumn{1}{C|}{\begin{tabular}[c]{@{}c@{}}\textbf{Decision Type}\\{(Matching Order)}\end{tabular}} & \begin{tabular}[C|]{@{}c@{}}\textbf{\# of Target Classes} \\ (\# of labels per class)\end{tabular} & \multicolumn{1}{C|}{\begin{tabular}[c]{@{}c@{}}\textbf{Branch Conditions}\\(Class lists or value ranges are separated by semicolons.)\end{tabular}}\\ \hline 
\endfirsthead
\caption{The statistics of 77 applications in empirical study (Continued).}\\
\hline 
\multicolumn{1}{|C|}{\textbf{Application}} & \multicolumn{1}{C|}{\begin{tabular}[c]{@{}c@{}}\textbf{Decision Type}\\{(Matching Order)}\end{tabular}} & \begin{tabular}[C|]{@{}c@{}}\textbf{\# of Target Classes} \\ (\# of labels per class)\end{tabular} & \multicolumn{1}{C|}{\begin{tabular}[c]{@{}c@{}}\textbf{Branch Conditions}\\(Class lists or value ranges are separated by semicolons.)\end{tabular}}\\ \hline 
\endhead
\hline
\multicolumn{4}{r}{\footnotesize( To be continued )}
\endfoot
\hline
\endlastfoot

\multicolumn{4}{|C|}{Image Multi-Label Classification (Google \code{label\_detection}, AWS \code{detect\_labels})} \\ \hline
\href{https://github.com/spaceqorgi/2019-iot-ai-workshop}{2019-iot-ai-workshop}                    &   Multi-Choice-App                        & \textbf{2} (7, 2)                & [Capuchin monkey, ...]; [Wildlife biologist, ...]                                                          \\
\href{https://github.com/Grusinator/Aander-ETL}{Aander-ETL}                               &   Multi-Choice-App                        & \textbf{3} (9, 6, 5)             & [Landmark, Sculpture, ...]; [Building, Estate, ...]; [Human, ...]                                      \\
\href{https://github.com/SmartAppUnipi/ArtGuide}{ArtGuide}                                 &   Multi-Choice-API                            & \textbf{2} (6, 3)                & [Painting, Picture frame, ...]; [Building, Architecture, ...]                                              \\
\href{https://github.com/ShruthiAthikam/AWS_CloudComputing}{AWS\_CloudComputing}                      &  Multi-Select                           & \textbf{2} (1, 1)                & [Hot dog]; [Food]                                                                                          \\
\href{https://github.com/amruthasingh/DoorWatch}{DoorWatch}                   & True-False                               & \textbf{1} (6)                   & [Clothing, Person, Human, Furniture, Child, Man]                                                               \\
\href{https://github.com/Vickyilango/AWSRekognition}{AWSRekognition}                           &  Multi-Select                           & \textbf{2} (3, 3)                & [Person, People, Human]; [Art, Drawing, Sketch]                                                            \\
\href{https://github.com/hsunchi/GraduateProject}{GraduateProject}                             & True-False                               & \textbf{1} (5)                   & [Orator, Professor, Projection Screen, ...]                                                          \\
\href{https://github.com/Sarveshtg/-Voice-Assistant-for-Visually-Impaired}{Voice-Assistant}                       &  Multi-Select                           & \textbf{3} (5, 3, 1)          & [Highway, Lane, ...]; [Car, ...]; [Classroom]                                 \\
\href{https://github.com/julian-gamboa-ensino/callforcode}{callforcode}                              & True-False                               & \textbf{1} (5)                   & [Water, Waste, Bottle, Plastic, Pollution]                                                                     \\
\href{https://github.com/J-eld/AWS-Rekognition-Car-Image-search}{Car-Image-search}                        & True-False                               & \textbf{1} (6)                   & [Sedan, Mini SUV, Coupe utility, Truck, Van, Convertible]                                                      \\
\href{https://github.com/sanjay417/cloudComputing_project2}{cloudComputing\_project2}                          &   Multi-Choice-API                            & \textbf{3} (1, 3, 1)             & [Person]; [Dog, Cat, Mammal]; [Flower]                                                                 \\
\href{https://github.com/bhavani-goruganthu/CSC847_GAE_Proj2_VisionAPI}{CSC847\_GAE\_Proj2}                             &   Multi-Choice-API                            & \textbf{3} (2, 2, 1)             & [Mammal, Livestock]; [Human, People]; [Flower]                                                         \\
\href{https://github.com/cmonyeba/cutiehack}{cutiehack}                                &  Multi-Select                           & \textbf{2} (1, 3)                & [Banana]; [Lemon, Citrus fruit, Apple]                                                                     \\
\href{https://github.com/cloudwaysX/CycleGAN-tensorflow_pixie}{CycleGAN-tensorflow\_pixie}                          &   Multi-Choice-API                            & \textbf{3} (1, 6, 4)             & [Food]; [Girl, Boy, Man, ...]; [Room, Living room, House, ...]                                         \\
\href{https://github.com/CalvinKrist/DisasterRelief}{DisasterRelief}                           & True-False                               & \textbf{1} (8)                   & [Hurricane, Flood, Tornado, Landslide, Earthquake, Volcano, ...]                                               \\
\href{https://github.com/sun624/Dogecoin_musk}{Dogecoin\_musk}                           & True-False                               & \textbf{1} (4)                   & [Dog, Mammal, Carnivore, Wolf]                                                                                 \\
\href{https://github.com/desendoo/flaskAPI}{flaskAPI}                                & True-False                               & \textbf{1} (3)                   & [Food, Recipe, Ingredient]                                                                                     \\
\href{https://github.com/lyl0602/Cloud-based-automatic-food-assessment-system}{food-assessment-system}                  &  Multi-Choice-API                           & \textbf{5} (35, 22, 54, 4, 6)    & [Dessert, ...]; [Grilling, ...]; [Strawberries, ...]; [Cigarette, ...]; ...        \\
\href{https://github.com/pruthu-vi/CV-Project2/}{Foodier}                                  &  Multi-Select                           & \textbf{2} (13, 1)               & [Building, Logo, Menu, Person, Vehicle, People, ...]; [Food]                                               \\
\href{https://github.com/neeltron/Hack-At-Home-II}{Hack-At-Home-II}                          &   Multi-Choice-API                            & \textbf{2} (3, 3)                & [Food, Junk food, Plastic]; [Drinkware, Wood, Metal]                                                       \\
\href{https://github.com/matthew-chu/heapsortcypher}{HeapSortCypher}                           &   Multi-Choice-API                            & \textbf{3} (8, 5, 11)            & [Food, Food grain, ...]; [Clothing, Shirt, ...]; [Paper bag, ...]                                      \\
\href{https://github.com/laksh22/IngredientPrediction}{IngredientPrediction}                                &  Multi-Select                        & \textbf{3} (1, 1, 1)             & [Spaghetti]; [Bean]; [Naan]                                                                        \\
\href{https://github.com/mattheweis/FESMKMITL}{FESMKMITL}                                & True-False                               & \textbf{1} (1)                   & [Face]                                                         \\
\href{https://github.com/dafna1228/milab}{milab}                                  &   Multi-Choice-App                        & \textbf{3} (1, 1, 1)             & [Sign]; [Nature]; [Car]                                                                                \\
\href{https://github.com/gdsc-ssu/bird-sweeper}{BirdSwe}                         &   Multi-Choice-API                            & \textbf{1} (5)                & [Smoke, Bird, ...]                                                       \\
\href{https://github.com/BONITA-KWKim/ai-server-proto}{ai-server-proto}                      &   Multi-Choice-API                            & \textbf{3} (3, 14)                   & [Eye, Eyeball, Eyes]; [Landmark, Sculpture, Monument, ...]                                                                                                \\
\href{https://github.com/Flowmot1on/Phoenix}{Pheonix}                                  & True-False                               & \textbf{1} (1)                   & [Fire]                                                                                                         \\
\href{https://github.com/nina-mir/photo_book_google_app_engine}{photo\_book}                              &   Multi-Choice-API                            & \textbf{3} (10, 10, 2)           & [Mammal, Bird, Insect, ...]; [Skin, Lip, ...]; [Flower, Plant]                                         \\
\href{https://github.com/siwasu17/plant-watcher/}{Plant-Watcher}                            & True-False                               & \textbf{1} (5)                   & [Plant, Flowerpot, Houseplant, Bonsai, Wood]                                                                   \\
\href{https://github.com/hamzaish/RecycleBot}{RecBot}                               &   Multi-Choice-App                        & \textbf{2} (11, 8)               & [Tin, Paper, Magazine, Carton, ...]; [Food, Bread, Pizza, ...]                                             \\
\href{https://github.com/adrian-willi/roblab-hslu}{roblab-hslu}                              & True-False                               & \textbf{1} (7)                   & [Raincoat, Coat, Jacket, T-shirt, Trousers, Jeans, Shorts]                                                     \\
\href{https://github.com/kyu929/senior-project/}{senior-project}                           &   Multi-Choice-API                            & \textbf{3} (2, 3, 1)             & [Landscape, Landmark]; [Self-portrait, Portrait, ...]; [Flower]                                        \\
\href{https://github.com/ertheosiswadi/smart_can}{smart-can}                                & True-False                               & \textbf{1} (9)                   & [Paper, Bottle, Plastic, Container, Tin can, Glass, ...]                                                       \\
\href{https://github.com/cod-r/smart-trash-bin}{smart-trash-bin}                        &   Multi-Choice-API                            & \textbf{2} (14, 5)               & [Aviator sunglass, Beer glass, ...]; [Plastic arts, ...]                                                   \\
\href{https://github.com/yuvsc/smartHamper}{smartHamper}                              &   Multi-Choice-API                            & \textbf{3} (7, 4, 3)             & [Shirt, T-shirt, ...]; [Trousers, Denim, ...]; [Brand, Text, ...]                                      \\
\href{https://github.com/cmfabregas/StudySpaceAvailability}{StudySpaceAvailability}                   & True-False                               & \textbf{1} (4)                   & [Hardware, Power Drill, Drill, Electronics]                                                                    \\
\href{https://github.com/The-Coding-Kid/888hacks-flask}{The-Coding-Kid}                           &  Multi-Select                           & \textbf{6} (9, 6, 9, 3, 6, 3)    & [Noodle, ...]; [Meat, ...]; [Produce, ...]; [Fruit, ...]; [Milk, ...]; ... \\
\href{https://github.com/pankeshpatel/tinyml-computer-vision}{Tinyml}                  &  Multi-Select                           & \textbf{3} (4, 6, 5)             & [Car, Truck, ...]; [Gun, Weapon Violence, ...]; [Cat, Dog, ...]                                        \\
\href{https://github.com/yifei-tang/UofTHacksBackend}{UofTHacksBackend}                         &   Multi-Choice-API                            & \textbf{4} (3, 3, 7, 4)          & [T-shirt, ...]; [Outerwear, ...]; [Pants, ...]; [Footwear, ...]                                    \\
\href{https://github.com/gnawcire/garbage-sort}{garbage-sort}                          &   Multi-Choice-API                           & \textbf{2} (1, 20)    & [Food];[Metal, ...];          \\
\hline \hline
\multicolumn{4}{|C|}{Image Object Detection (Google \code{object\_localization})}\\\hline
\href{https://github.com/tjestes/equipment-detection-poc}{equipment-detection-poc}                  &  Multi-Select                          & \textbf{1} (1)                   & [Shoe]                                                                                                         \\
\href{https://github.com/nlonberg/flood-depths/}{flood\_depths}                            &  Multi-Select                           & \textbf{1} (5)                   & [Car, Van, Truck, Boat, Toy vehicle]                                                                           \\
\href{https://github.com/qwerty10w/SBHacks2021/}{SBHacks2021}                              &  Multi-Select                          & \textbf{1} (1)                   & [Person]                                                                                                       \\
\href{https://github.com/arosloff/SeeFarBeyond}{SeeFarBeyond}                             &  Multi-Select  & \textbf{1} (2)  & [Spoon, Coin]                                                                                                  \\
\href{https://github.com/thy0602/shecodes-hack}{shecodes-hack}                            &  Multi-Select                           & \textbf{1} (2)                   & [Dress, Top]                                                                                                   \\
\href{https://github.com/renilJoseph/SunHacks2019/}{SunHacks2019}                             &  Multi-Select                          & \textbf{1} (3)                   & [Person, Chair, Table]                                                                                         \\
\href{https://github.com/rlathgml/thgml/}{thgml}                                    &  Multi-Select                          & \textbf{1} (7)                   & [Pizza, Food, Sushi, Baked goods, Snack, Cake, Dessert]                                                        \\
\href{https://github.com/sarvesh-tech/Verlan/}{Verlan}                                   &  Multi-Select                        & \textbf{1} (2)                   & [Dog, Animal]                                                                                                  \\
\hline \hline
\multicolumn{4}{|C|}{Text Sentiment Classification (Google \code{sentiment\_detection})}\\\hline 
\href{https://github.com/OkapalDominic/animal_analysis}{animal-analysis}                          &   Multi-Choice-API                            & \textbf{4} (1, 1, 1, 1)          & [0.5, 1]; [0, 0.5]; [-0.5, 0]; [-1, -0.5]                                                          \\
\href{https://github.com/kmzjy110/calhacksv2}{calhacksv2}                               &   Multi-Choice-API                            & \textbf{6} (1, 1, 1, 1, 1, 1)    & [0.5, 1]; [0.5, 1]; [0.1, 0.5]; [-0.1, 0.1]; [-0.5, -0.1]; [-1, -0.5]                      \\
\href{https://github.com/steventhan/carbon-hack-sentiment}{carbon\_hack\_sentiment}                  &   Multi-Choice-API                            & \textbf{3} (1, 1, 1)             & [0.3333, 1]; [-0.3333, 0.3333]; [-1, -0.3333]                                                          \\
\href{https://github.com/Martincu-Petru/Cloud-Computing}{FoodDelivery}                         &   Multi-Choice-API                            & \textbf{3} (1, 1, 1)             & [0.6, 1]; [0.3, 0.6]; [-1, 0.3]                                                                        \\
\href{https://github.com/ryanphennessy/devfest}{devfest}                                   &   Multi-Choice-API                            & \textbf{4} (1, 1, 1, 1)          & [0.6, 1]; [0.4, 0.6]; [0.2, 0.4]; [-1, 0.2]                                                        \\
\href{https://github.com/ChainZeeLi/EC601_twitter_keyword}{EC601\_twitter\_keyword}                  &   Multi-Choice-API                            & \textbf{3} (1, 1, 1)             & [0.25, 1]; [-0.25, 0.25]; [0.25, 1]                                                                    \\
\href{https://github.com/Dacs95/ElectionSentimentAnalysis}{ElectionSentimentAnalysis}                 &   Multi-Choice-API                            & \textbf{3} (1, 1, 1)             & [0.05, 1]; [0, 0.05]; [-1, 0]                                                                          \\
\href{https://github.com/jtkrumlauf/Hapi}{Hapi}                                        &   Multi-Choice-API                            & \textbf{2} (1, 1)                & [-1, 0]; [0, 1]                                                                                            \\
\href{https://github.com/beekarthik/JournalBot}{JournalBot}                               &   Multi-Choice-API                            & \textbf{3} (1, 1, 1)             & [0.5, 1]; [0, 0.5]; [-1, 0]                                                                            \\
\href{https://github.com/whtai/Mind-Reading-Journal/}{Mind\_Reading\_Journal}                   &   Multi-Choice-API                            & \textbf{4} (1, 1, 1, 1)          & [0.15, 1]; [0.1, 0.15]; [-0.15, 0.1]; [-1, -0.15]                                                  \\
\href{https://github.com/Mrkr1sher/Sarcatchtic-MakeSPP19}{Sarcatchtic-MakeSPP19}                  &   Multi-Choice-API                            & \textbf{2} (1, 1)                & [-0.5, 1]; [-1, -0.5]                                                                                      \\
\href{https://github.com/nicholasadamou/stockmine}{stockmine}                                &   Multi-Choice-API                            & \textbf{2} (1, 1)                & [-1, 0]; [0, 1]                                                                                            \\
\href{https://github.com/KijanaG/Tone}{Tone}                                     &   Multi-Choice-API                            & \textbf{3} (1, 1, 1)             & [-1, -0.5]; [-0.5, 0.5]; [0.5, 1]                                                                      \\
\href{https://github.com/nixin72/UOttaHack-2019}{UOttaHack\_2019}                          &   Multi-Choice-API                            & \textbf{3} (1, 1, 1)             & [0.25, 1]; [-0.25, 0.25]; [-1, -0.25]                                                                  \\
\hline \hline
\multicolumn{4}{|C|}{Text Entity Detection (Google \code{entity\_detection})}\\\hline
\href{https://github.com/Jhuynh760/GeoScholar}{GeoScholar}                               & True-False                               & \textbf{1} (1)                & [LOC]                                                                                   \\
\href{https://github.com/mihirKachroo/HackThe6ix}{HackThe6ix}                               &   Multi-Choice-API                            & \textbf{7} (1, 1, 1, 1, 1, 1, 1) & [PERSON]; [LOC]; [ADD]; [NUM]; [DATE]; [PRICE]; [ORG]             \\
\href{https://github.com/dev5151/Klassroom}{Klassroom}                                &   Multi-Choice-API                            & \textbf{2} (2, 2)                & [PERSON, PROPER]; [LOC, ORG]                                                                 \\
\href{https://github.com/da1234/newsChronicle/}{newsChronicle}                            & True-False                               & \textbf{1} (1)                   & [OTHER]                                                                                                        \\
\href{https://github.com/larry852/ocr-contratos/}{ocr-contratos}                           & True-False                               & \textbf{1} (1)                   & [NUM]                                                                                                          \\
\href{https://github.com/AllegraChen/uofthacks6}{uofthacks6}                               & True-False                               & \textbf{1} (1)                   & [OTHER]                                                                                                        \\
\hline \hline
\multicolumn{4}{|C|}{Text Topic Classification (Google \code{text\_classify})} \\ \hline
\href{https://github.com/Shrinjay/DMnMD}{DMnMD}                                    & True-False                               & \textbf{1} (1)                   & [Health]                                                                                                       \\
\href{https://github.com/saheedandrew/HLPFL}{HLPFL}                                    & True-False                               & \textbf{1} (8)                   & [Public Safety, Law \& Government, Emergency Services, News, ...]                                              \\
\href{https://github.com/SaiManukonda/MirrorDashboard}{MirrorDashboard}                          & True-False                               & \textbf{1} (7)                   & [Jobs \& Education, Law \& Government, News, ...]                                                              \\
\href{https://github.com/GalenWong/noteScript}{noteScript}                               & True-False                               & \textbf{1} (1)                   & [Food]                                                                                                         \\
\href{https://github.com/dwang/pennapps-2019f}{pennapps\_2019f}                          & True-False                               & \textbf{1} (2)                   & [News/Politics, Investing]                                                                                     \\
\href{https://github.com/jcavejr/soap}{soap}                                     &   Multi-Choice-API                            & \textbf{2} (2, 2)                & [Sensitive Subjects, ...]; [Discrimination \& Identity Relations, ...]                                     \\
\href{https://github.com/Samvit123/SocialEyes-MakeSPP2018}{SocialEyes}                               &   Multi-Choice-API                            & \textbf{2} (2, 1)                & [people \& society, sensitive subjects]; [adult]                                                           \\
\href{https://github.com/gpesma/Twitter-Mining-GAE}{Twitter\_Mining\_GAE}                     & True-False                               & \textbf{1} (1)                   & [Sentitive]                                                                                                    \\
\href{https://github.com/JoosepAlviste/vfriendo}{vfriendo}                                 & True-False                               & \textbf{1} (1)                   & [Restaurants]                                                                             
\end{tabularx}

\twocolumn 

\section{Loss function for other \summaries}
\label{app:loss_appendix}

 \mypara{\term{True-False}}

\begin{align}
     L(y) &= \overbrace{ \textrm{Sigmoid}(\max_{\LabelId \in \TargetSet_{\hat{\ClassId}}}(\y) - \theta)
}^{\substack{\text{\sf \footnotesize {Penalize {\bf Type-1} Critical Errors}} 
}} \nonumber \\&+
\overbrace{ \textrm{Sigmoid}(\theta-\max_{\LabelId \in \TargetSet_{\ClassId}}(\y))
}^{\substack{\text{\sf \footnotesize {Penalize {\bf Type-1} Critical Errors}} 
}} 
\end{align}

\mypara{\term{Multi-Select}}
\begin{equation}
\begin{aligned}
     L(y) &= \overbrace{ \sum_{\ClassId \in \hat{T}}\textrm{Sigmoid}(\theta - \max_{\LabelId \in \TargetSet{\ClassId}} \y[\LabelId] )
}^{\substack{\text{\sf \footnotesize {Penalize {\bf Type-1} Critical Errors}} 
}} \\
&+\overbrace{ \sum_{\ClassId \in \cup_{\ClassId} \TargetSet_{\ClassId} \setminus \hat{T}}\textrm{Sigmoid}(\max_{\LabelId \in \TargetSet{\ClassId}} \y[\LabelId] - \theta)
}^{\substack{\text{\sf \footnotesize {Penalize {\bf Type-3} Critical Errors}} 
}} 
\end{aligned}
\end{equation}

\mypara{\term{Multi-Choice API-order}}
Here we explain why this loss function captures the critical errors:
\begin{packeditemize}
    \item A Type-1 error occurs, if (1) the correct \target is matched, thus at least one of its labels has a score above the confidence threshold ($\max_{\LabelId \in \TargetSet_{\hat{\ClassId}}} \y[\LabelId] \geq \theta$), and (2) it is matched after the EOD because all of the labels belonging to the correct \target have scores below the maximum score of labels in the incorrect \targets. 
    \item A Type-2 error occurs if the maximum score for labels in a correct \target falls below threshold $\theta$, thus it is never matched (before or after EOD).
    \item A Type-3 error occurs if any labels belonging ($\max_{\LabelId \notin \TargetSet_{\hat{\ClassId}}} \y[\LabelId]$) to  incorrect \targets appears before labels in the correct \target. 
    
\end{packeditemize}

\begin{equation}
\begin{aligned}
     L(\y) &=   
  \overbrace{ \textrm{Sigmoid}\left(\max_{\LabelId \in \cup_{\ClassId \neq \hat{\ClassId} }\TargetSet_{\ClassId}}\y[\LabelId] - \max_{\LabelId \in \TargetSet_{\hat{\ClassId}}} \y[\LabelId]\right) }^{\substack{\text{\sf \footnotesize {{\bf Type-1} Critical Errors}} }} \\
  &+
  \overbrace{ \textrm{Sigmoid}\left(\max_{\LabelId \in \cup_{\ClassId \neq \hat{\ClassId} }\TargetSet_{\ClassId}}\y[\LabelId] - \theta\right)}^{\substack{\text{\sf \footnotesize {{\bf Type-2} Critical Errors}} }}\\
  &+
  \overbrace{ \sum_{\ClassId \neq \hat{\ClassId}}\textrm{Sigmoid}\left( \max_{\LabelId \in \TargetSet_{\ClassId}}\y[\LabelId] 
- \max_{\LabelId \in \TargetSet_{\hat{\ClassId}}}\y[\LabelId] \right)}^{\substack{\text{\sf \footnotesize {{\bf Type-3} Critical Errors}} }}
  \label{eq:loss_app2}
\end{aligned}
\end{equation}

\mypara{\term{Value ranges}} As for APIs that output a score $\y$ to describe the input, applications typically define several value ranges as target classes to make decisions, where the lower bound of the $\ClassId^{th}$ \target is denoted as $l_\ClassId$ and the upper bound of the $\ClassId^{th}$ \target is denoted as $u_\ClassId$. 

\begin{equation}
\begin{aligned}
L(y_i) &= 
\overbrace{ \textrm{Sigmoid}\left(\y - u_{\hat\ClassId} \right) +
\textrm{Sigmoid}\left(l_{\hat\ClassId}- \y \right)  }^{\substack{\text{\sf \footnotesize {{\bf Type-1} Critical Errors}} }} \\&+
\overbrace{ \sum_{\ClassId \neq \hat{\ClassId}} \textrm{Sigmoid}(u_{\ClassId} - \y) + \textrm{Sigmoid}(\y - l_{\ClassId} ) }^{\substack{\text{\sf \footnotesize {{\bf Type-3} Critical Errors}} }} 
 \label{eq:value}
\end{aligned}
\end{equation}
where $\hat{\ClassId}$ is the index of the correct \target. 
A Type-1 error occurs (\ie a correct \target is matched after EOD) when the output score $\y$ exceeds the upper bound of the ground-truth value range ($u_{\ClassId}$), or falls below the lower bound of the ground-truth value range ($l_{\ClassId}$).
A Type-3 error occurs when the upper bound of an incorrect value range exceeds $\y$ \emph{and} its lower bound falls below $\y$, leading it to be selected. 
Type-2 errors are absent in this application because all the \targets span the whole output range, thus a \target must be matched.

\newpage
\edit{
\section{Setup of \prespec}
\label{appendix:appendix_complete}
Here, we describe in detail how we construct the label categories to support the scheme of \prespec, which is one of the schemes in comparison with \tool (Section~\ref{sec:evaluation}).

\paragraph{Image-classification} Image-classification APIs typically contains many thousands of labels without providing their categorization or hierarchy. 
Therefore, we create categories leveraging the Wikidata knowledge graph~\cite{wikidata}, a widely used knowledge graph database that has been referred to during the creation of many popular ML training datasets~\cite{haller2022analysis, openimages, DBLP:journals/corr/JundEAB17}.
In this knowledge graph, each node is a named entity, covering \textit{all} the labels used in popular image-classification APIs~\cite{google-cloud,amazon-ai,ms-azure-image-tag}, and each edge represents a relationship between two entities (e.g., ``subclass of'', ``different from'', ``said to be the same as''). 

Extracting ``subclass of'' edges in Wikidata knowledge graph, we get a directed acyclic graph (DAG) of label hierarchy.
We believe it offers a principled foundation to create label categories based on two observations: 
(1) If an entity/node $e$ is reachable from another entity/node $e'$ through several subclass-of edges, $e'$ is also covered by the \textit{category} of $e$ (e.g., entity ``motor vehicle'' is directly connected to ``land vehicle'', entity ``land vehicle'' is directly connected to ``vehicle'', so ``motor vehicle'' is also covered by the ``vehicle'' category);
(2) The distance, measured in the number of subclass-of edges, between an entity/node and the DAG root indicates the specificity of the concept behind this node, with shorter distance representing coarser-grained categories.
We will refer to a node that is $k$ edges away from the root as a Level-$k$ node.

Based on these observations, we formally define a set of categories $C_k$ for all the image-classification labels ${L}$ at a specificity level $k$ as follows:
$C_k$ is the minimum set of Level-$k$ nodes such that every label $l \in L$ is covered by at least one category node $c_k \in C_k$. 
We could categorize all the applications into single-category and multi-category applications using any level of specificity settings. 
In this paper, we adopt Level-2 specificity setting, since the number of single-category applications drops a lot when moving from Level-2 to Level-3, indicating Level-3 categories may be too fine-grained. 

Under Level-2, we set up 35 categorized models that cover all the image-classification labels.  
Six of them are used by applications in our benchmark, including \textit{natural object}, \textit{temporal entity}, \textit{artificial entity}, \textit{system}, \textit{phenomenon}, and \textit{continuant}. 
With this categorization, 27 of the 40 image-classification applications are single-category, and the rest 13 applications are multi-category.

\paragraph{Object-detection} Similar as image-classification API, every object-detection API label also corresponds to an entity node in the Wikidata knowledge graph. Therefore, we use the same methodology and the same specificity Level-2 to define categories for object detection labels.

Seven categorized models are set up to cover all object-detection labels. 
The object-detection labels used by 8 object-detection benchmark applications belong to 3 categories: \emph{natural object}, \emph{artificial entity}, and \emph{system}.
Under this setting, there will be 7 single-category applications, and 1 multi-category applications. 

\paragraph{Text-classification} The Google text-classification API~\cite{google-cloud} offers the hierarchy tree of all its labels. We simply follow their categorization and get 15 categories to covering all text-classification labels. Nine categories are used by applications in our benchmark, including \emph{business \& industrial}, \emph{people \& society}, \emph{health}, \emph{food \& drink}, \emph{jobs \& education}, \emph{news}, \emph{sensitive subjects}, \emph{adult}, and \emph{law \& government}. 
Under this categorization, there are 5 single-category text-classification applications, and 4 multi-category text-classification applications.

\paragraph{Other types of applications} There are two other types of applications in our benchmark that are not suitable for designing pre-specialized models: sentiment analysis and named entity recognition. 

For sentiment analysis API, the corresponding applications typically define several value-ranges and determine which range the API output (a sentiment score) falls into. Since the API output is a floating point number, there are infinite ways of defining value-ranges. Therefore, it is impracticable to create pre-categorized models. 

For named entity recognition API, it only has 6 labels: {\em person, location, organization, number, date}, and \emph{misc}~\cite{google-cloud}. They are already high-level categories. There is no need to create pre-categorized models for each category.
}





\newpage
\onecolumn
\section{Results of other applications}
\label{appendix:appendix_otherapp}
\begin{figure}[h]
\centering
    \includegraphics[width=0.5\textwidth]{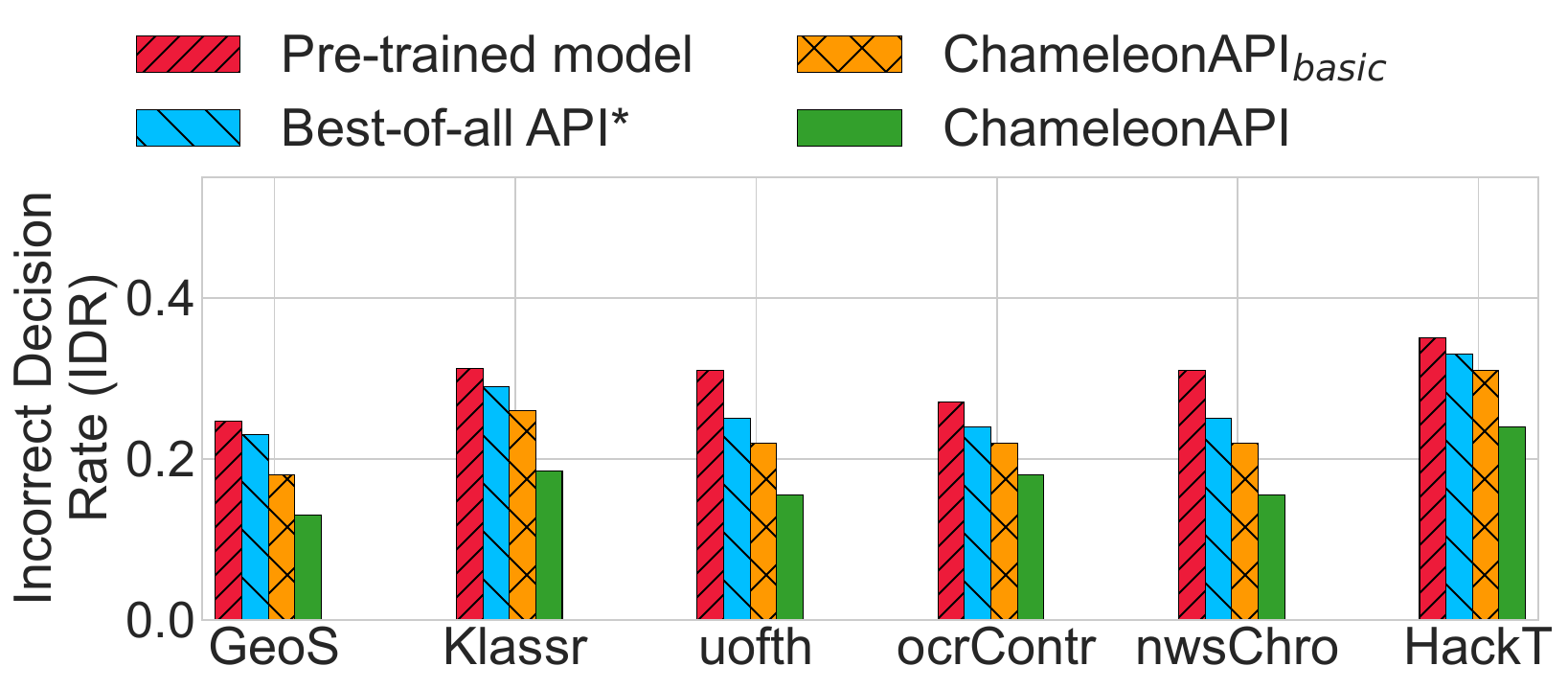} 
    \caption{\edit{Results on entity-recognition applications.}}
    \label{fig:entity}
\end{figure}

\begin{figure*}[h]
\centering
   \includegraphics[width=0.8\textwidth]{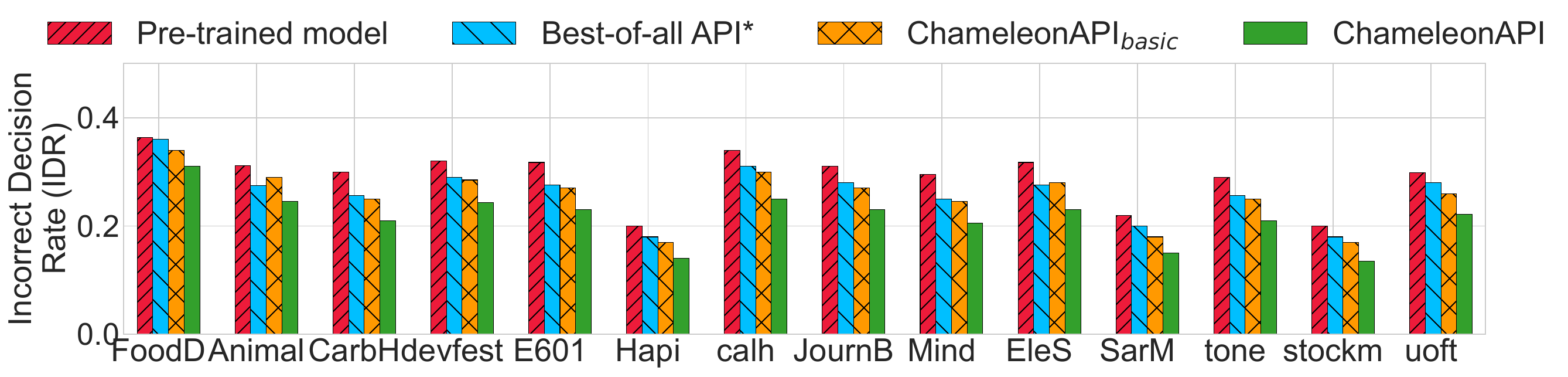} 
        \caption{\edit{Results on sentiment-analysis applications.} }
        \label{fig:sentiment_analysis}
\end{figure*}

As mentioned in \S\ref{sec:eval_setup}, the results of the 20 applications
that involve sentiment analysis and entity recognition were not included
in the evaluation section. Their results are shown in 
Figure~\ref{fig:entity} and~\ref{fig:sentiment_analysis}.
As we can see, the advantage of \tool is consistent across these applications, similar to what we presented in\S\ref{sec:evaluation}.
Note that, the scheme of \prespec does not apply to applications that 
involve these two types of ML tasks, and hence is not included in Figure 
\ref{fig:entity} and \ref{fig:sentiment_analysis}.

\end{appendices}